\pdfoutput=1
\documentclass[a4paper, 12 pt]{article}
\usepackage{lipsum}
\usepackage{graphicx,subcaption}
\usepackage{float}
\usepackage{bm}
\usepackage{mathtools}
\usepackage{caption}
\usepackage{wrapfig}
\usepackage{pdfpages}
\usepackage{epsfig}
\usepackage{epstopdf}
\usepackage{listings}
\usepackage{verbatim}
\usepackage{amsmath,amsfonts,amssymb}
\usepackage{xcolor}
\usepackage{array}
\newcolumntype{P}[1]{>{\centering\arraybackslash}p{#1}}

\usepackage{hyperref}
\hypersetup{
    colorlinks=true,
    linkcolor=blue,
    filecolor=magenta,      
    urlcolor=cyan,
    citecolor=cyan
}

\date{} %

\usepackage[numbers]{natbib}

\begin{document}



\title{Noise-induced synchronization in the Kuramoto model on finite 2D lattice}                      

\author{Mrinal Sarkar}
\date{\textit{Department of Physics,
Indian Institute of Technology Madras, Chennai-600036, India.}\\
Email: \texttt{mrinalsarkar92@gmail.com}\\~\\
    {(Dated: \today)}
}


\maketitle




\begin{abstract}
We consider the celebrated Kuramoto model with nearest neighbour interactions, arranged on a two-dimensional square lattice in presence of two kinds of noise - annealed and quenched.
	We focus on both the steady state and relaxation dynamics of the model. The 
	bare model with annealed noise on finite $2D$ lattice, in the stationary state, exhibits a crossover from synchronization to desynchronization as noise strength varies. Finite-size scaling (FSS) analysis reveals that, in the thermodynamic limit, this crossover becomes a true phase transition, which is Kosterlitz-Thouless ($KT$)-type analogous to that of $2D$ $XY$ model, where the system makes a transition from a state with quasi-long range order (critically ordered phase) at low noise-strength to complete disorder at high noise-strength. 
	The critical noise-strength as well as the critical exponents associated with the transition are estimated using FSS theory. 
	On the other hand, when the noise is quenched,
	it does not show any kind of synchronization-desynchronization phase transition in the thermodynamic limit. 
    But we do observe a crossover from low noise-strength synchronization to high noise-strength desynchronization in the system of finite-size. We analyze the crossover phenomena through the linear stability of the stationary state solutions and obtain the crossover noise-strength from the onset of local instability of the unsynchronized one. The inverse crossover noise-strength is found to scale with the system-size logarithmically. The relaxation dynamics also differs for these two types of noise. In case of annealed noise, the system, in the critically ordered phase, exhibits algebraic relaxation which is described by the phenomenological Edwards-Wilkinson (EW) model of growing surface, yielding the same dynamic exponent $ z =2$. In disordered phase, the system shows an exponential decay. On the contrary, the system with quenched noise, as opposed to the annealed one, always relaxes to the stationary state exponentially. Both the system-size and noise-strength dependency of the average relaxation time in the synchronized regime are also investigated.
\end{abstract}





\section{Introduction}
Complex systems consisting of large population of coupled oscillators show a variety of rich emergent behaviours. Collective synchronization is one of the most fascinating ones
\cite{pikovsky2003synchronization,strogatz2004sync,gupta2018statistical}. This phenomenon is ubiquitous in different disciplines of science, including physics, chemistry, biology, social science and so on. Few typical examples in physics where collective synchronization emerges are arrays of Josephson junctions \cite{wiesenfeld1996synchronization,vlasov2013synchronization,ovchinnikov2013networks}, laser arrays \cite{silber1993stability,deshazer2001detecting} etc. In biological systems such as yeast cell suspensions \cite{bier2000yeast}, cardiac pacemaker cells \cite{winfree2001geometry}, neural networks \cite{mirollo1990synchronization,varela2001brainweb} this phenomenon is observed. The list is quite extensive. For more examples on synchronization, we refer the article \cite{acebron2005kuramoto}. The journey started long back, in the seventeenth century, when Huygens observed asynchronous behaviour of a couple of pendulum clocks hanging from a common support. 
But it took a long time to receive scientists' attention to the problem. Over the last few decades, only after the work of A.T. Winfree and Y. Kuramoto, it has been studied rigorously being motivated by these different biological and physical phenomena.

The Kuramoto model is the paradigmatic model to study synchronization in many-body interacting systems. The original model consists of a population of globally-coupled oscillators with distributed natural frequencies. This model, being simple and analytically tractable, has been studied in great details for a long time \cite{kuramoto1975international,kuramoto1986lect,strogatz2000kuramoto,hong2007entrainmentc}. It shows a large variety of synchronization patterns and has many applications in different contexts \cite{kuramoto2002coexistence,tinsley2012chimera,nkomo2013chimera,flovik2016describing,breakspear2010generative}.

However, in most real world systems, the nature of interactions is complex and thus the mean-field version does not hold always. To mimic realistic systems, many variations of original version exist such as Kuramoto model with local coupling \cite{sakaguchi1987local,strogatz1988phase,aoyagi1991frequency,bahiana1994order,rogers1996phase,marodi2002synchronization,hong2005collective,lee2010vortices}, Kuramoto model with frustration \cite{daido1987population,daido1992quasientrainment,cross1993pattern,ha2018emergence}, Kuramoto model with inertia \cite{tanaka1997first,hong1999inertia,gupta2014nonequilibrium}, Kuramoto model in presence of noise \cite{sakaguchi1988cooperative,arenas1994exact,bonilla1998time,hong1999noise,komarov2014synchronization,gupta2014kuramoto}, Kuramoto oscillators on different types of graphs \cite{hong2007finite,ji2014low,um2014nature}, Kuramoto model with time-delayed coupling \cite{yeung1999time,wu2018dynamics,dai2018interplay}, to name a few.  Here, in this communication, we consider one such variation, where the interaction is local. We choose a system of Kuramoto oscillators arranged in a two-dimensional square lattice, where they interact with their nearest neighbours only. There are many examples in nature with this type of interaction. But, such spatially extended locally coupled system, being analytically intractable possibly, is less explored. Although, few studies on phase synchronization in such model have been reported before, many questions still remain to be answered.

We study phase synchronization in the bare Kuramoto model in presence of annealed and quenched types of noise. For both kinds of noise, as the noise-strength varies, the system of finite size exhibits a crossover from synchronized state at low noise-strength to de-synchronized state at high noise-strength. Naturally the question arises: whether does this crossover from synchronization to de-synchronization happen in the thermodynamic limit or not? If yes, whether it remains as a crossover or becomes a true, singular phase transition.
To address these questions, we first study the linearized version of the model in finite system in the weak noise-strength regime. But, the linear theory, being unable to predict about the possibility of phase transition in the system, the full non-linear system is studied numerically. 
For the full non-linear system, in presence of annealed noise, we examine the finite-size scaling (FSS) behaviour of the order parameter and its dynamic fluctuation to characterize the crossover. Using FSS, we show that the system of 2D Kuramoto oscillators exhibits a true phase transition as the noise-strength (in terms of temperature) varies in the limit the number of oscillators $N \to \infty$.

A detailed investigation of the FSS behaviour shows that this transition, in the thermodynamic limit, is not an usual order-disorder transition. It is actually a transition from low-temperature critically ordered phase to high-temperature disordered phase. The critical noise-strength as well as the exponents associated with the transition are also obtained. The exponents are found to be temperature-dependent. The existence of temperature-dependent exponents indicates vanishing order parameter and infinite fluctuations (diverging correlation length) at all temperatures upto a critical value in the thermodynamic limit. The Binder's cumulant for different system sizes stay collapsed upto critical noise-strength, supporting the same. We also examine the behaviour of two-point correlation function in these two phases and calculate the exponent, which characterizes the power-law behaviour of the correlation in the critically ordered phase, at different temperatures. We believe this transition, as FSS theory predicts in the thermodynamic limit, is equivalent to the topological phase transition in the $2D$ $XY$ model. 

On the other hand, when the noise is quenched, although the system of finite size "crosses over" from synchronization to desynchronization as a function of quenched noise-strength, it does not do so in the limit of infinite system-size.
We investigate the system-size dependency of the crossover noise-strength, as calculated from the maxima of the fluctuations in the order parameter. We explain the synchronization crossover on a finite lattice via local stability of its stationary states. For a finite lattice, the synchronized solutions are locally neutrally stable while the unsynchronized one is locally unstable.
We believe that the local instability of the unsynchronized solution takes place due to the so-called "runaway" oscillators present in the system. The noise-strength, at which the instability of the unsynchronized solutions sets in, yields the crossover point. The crossover noise-strength obtained from stability analysis matches with that obtained from direct simulation within some error. The inverse crossover noise-strength depends logarithmically on the system-size which is consistent with the work by Lee \textit {et al.} \cite{lee2010vortices}.

Along with the stationary state dynamics, we study the relaxation dynamics as well, by studying temporal evolution of the order parameter. In the critically ordered phase, the system with annealed noise, relaxes algebraically and it belongs to the Edward-Wilkinson (EW) universality class yielding the same dynamic exponent $z =2$. The system of oscillators with quenched noise, on the other hand, follows an exponential law even in the synchronized phase. In this case, the average relaxation time is calculated and its variation with quenched noise-strength as well as system sizes is also investigated. On the other hand, in disordered phase, for both types of noise, the decay is exponential. 


The paper is organized as follows. In Section \ref{The Model}, we describe the model to be studied along with a summary of earlier works and our queries. In Section \ref{PT:annealed noise}, the stationary state dynamics of the model with annealed noise is discussed. Single oscillator distribution in finite system is obtained using the linearized version of the model. 
We investigate the FSS behaviour of different statistical quantities to study synchronization and unveil a true phase transition in the thermodynamic limit, which is $KT$-type, for the full non-linear system in Section \ref{PT:KT}.
In section \ref{PT:quenched noise}, we revisit the model with the quenched noise briefly and compare our results with the literature.
Next, we investigate the relaxation dynamics for both types of noise which is discussed in Section \ref{RD}. The paper ends with conclusions and future direction of our work. 
We establish the relationship of the model under study with the classical $XY$ model in Appendix \ref{appendix_KM_XY_equivalence}, followed by a brief discussion on $KT$ transition in the $2D$ $XY$ model in Appendix \ref{appendix_KT}.  Finally, the equivalence between the model with annealed noise at low noise-strength  and the Edwards-Wilkinson model is shown in Appendix \ref{appendix_EW}.

\section{The Model}
\label{The Model}
We consider the bare Kuramoto model with nearest neighbour interactions under the influence of two kinds of noise - annealed and quenched. The phase evolution equation of the $i$-{th} oscillator is then given by
\begin{equation}
\frac{{\rm d} \theta_{i}}{\rm dt} = \omega_{i} + K  \sum_{j \in nn_{i}}  \sin(\theta_{j}- \theta_{i}) + \eta_{i}(t),
\label{govn_eqn_1}
\end{equation}
where $nn$ represents that the sum is over nearest neighbours only, $\eta_{i}(t)$ is the \textit{annealed noise} term which is Gaussian white noise characterized by
\begin{equation}
\langle \eta_{i}(t) \rangle = 0    \qquad\mbox{and}\qquad  \langle \eta_{i}(t)\eta_{j}({t'}) \rangle = \Gamma \delta_{ij} \delta(t - {t}^{\prime}).
\end{equation}
Here, $\langle \cdot \rangle$ denotes averaging over noise realizations, $K (\geq 0)$ is the coupling strength, $\Gamma$ is the strength of the Gaussian noise. 
The $\omega_{i}$'s are  \textit{quenched noise} also drawn from a Gaussian distribution $ f(\omega) $ with mean zero and variance $\sigma^{2}$,
\begin{equation}
\langle \omega_{i} \rangle = 0    \qquad\mbox{and}\qquad  \langle \omega_{i}\omega_{j} \rangle = \sigma^{2}\delta_{ij}.
\end{equation}
To show explicit dependence of synchronization on these two types of noise-strengths, we rewrite the above evolution equation as follows:
\begin{equation}
\frac{{\rm d} \theta_{i}}{\rm dt} =  \sigma \omega_{i} + K  \sum_{j \in nn_{i}}  \sin(\theta_{j}- \theta_{i}) + \sqrt{\Gamma} \eta_{i}(t),
\label{govn_eqn_2}
\end{equation}
where both the distribution are now of zero mean and unit variance.
By proper rescaling of time, we can include the effect of coupling term in the other parameters and get
\begin{equation}
\frac{{\rm d} \theta_{i}}{\rm d\tilde{t}} =  \tilde{\sigma} \omega_{i} + \sum_{j \in nn_{i}}  \sin(\theta_{j}- \theta_{i}) + \tilde{g} \xi_{i}(\tilde{t}),
\label{govn_eqn_3}
\end{equation}
where
\begin{equation*}
  \tilde{t} \equiv Kt,\qquad\tilde{\sigma} \equiv \sigma/K,\qquad \tilde{g} \equiv \sqrt{\Gamma/K} \qquad\mbox{and}\qquad  \xi_{i}(\tilde{t}) \equiv \eta_{i}(t)/K.  
\end{equation*}
From now on, we will drop the tilde for simplicity of notations and study the dynamics of the system in terms of these two reduced dimensionless parameters, $\sigma$ and $g$.
\begin{equation}
\frac{{\rm d} \theta_{i}}{\rm dt} =  \sigma \omega_{i} + \sum_{j \in nn_{i}}  \sin(\theta_{j}- \theta_{i}) + g \xi_{i}(t)
\label{govn_eqn_4}
\end{equation}

Before we move on to study the system in details, we will have a more careful look into the model. The random variables in "quenched disorder" can also be thought of as intrinsic frequencies of the oscillators, and the "annealed" white noise can be assumed as the fluctuations in the system. So, as a whole, the model can be visualized as locally coupled Kuramoto model with natural frequencies drawn from a Gaussian distribution in presence of stochastic force, characterized by Gaussian white noise. Stochasticity is an inevitable characteristic of the phenomena observed in nature. The source of stochasticity may be different in different contexts. It may be intrinsic in some systems, e.g., in biological systems where the natural frequencies of the oscillators may have a fluctuating part. In thermodynamic systems, it arises from the thermal fluctuation present in the system.

The model described above has two important aspects corresponding to the two limiting cases: (a) the limit $g \to 0$ and (b) the limit $\sigma \to 0$.
 \begin{description}
  \item[$\bullet$ Case-1:] The limit $g \to 0$ corresponds to Kuramoto model with distributed natural frequencies in absence of gaussian white noise. So this is a non-linear $\it dynamical$ system which relaxes to a non-equilibrium stationary state at long time. 
  \item[$\bullet$ Case-2:] The limit $\sigma \to 0$ corresponds to Kuramoto model with identical oscillators in presence of gaussian white noise. The dynamics reduces to that of a $\it statistical$ system in contact with a heat bath the long time dynamics of which is governed by equilibrium statistical mechanics. 
  
\end{description}
 
 
\subsection{Previous works and our motivation}
The locally coupled Kuramoto model on different spatial dimensions with unimodal frequency distribution, in the form of quenched disorder, has already been studied. We summarize the results known for this system. Strogatz and Mirollo \cite{strogatz1988phase} proved analytically that no entrainment transition is possible in the thermodynamic limit in locally coupled one dimensional system. In fact, this is true for any spatial dimensions. That means, the lower critical dimension for phase-locking in the thermodynamic limit is infinite. However, in higher dimensions, clustering may happen but in such case, if large clusters of size $\mathcal{O}(N)$ ($N$ being the system size) exist, they must have sponge-like geometry i.e. the clusters are riddled with holes, which correspond to un-synchronized oscillators. But it does not rule out the possibility of crossover phenomena in finite systems. When the system size is finite, the dynamics does show a crossover from synchronization to de-synchronization as the coupling strength is varied. Hong \textit{et al.} \cite{hong2005collective} studied the possibility of phase transition for the full non-linear system in different spatial dimensions including $2D$. Lee \textit{et al.} \cite{lee2010vortices} showed how the entrainment crossover takes place in a two-dimensional lattice via the stability of vortex-antivortex pairs formed in the phase-field of the oscillators. For small system sizes, the crossover coupling-strength shows logarithmic dependence on it.

 All the previous studies on synchronization are only for the case of quenched noise (the limit $g \to 0$). Naturally, we ask the following question: what happens to its synchronization dynamics when the system is subjected to external random force which is annealed in nature. We study synchronization in finite systems first and investigate whether or not there is a possibility of phase transition in such a system in the thermodynamic limit.

In this communication, we consider a  system where the Kuramoto oscillators are arranged on a two-dimensional square lattice and each of them follows the evolution equation (\ref{govn_eqn_4}). We assume periodic boundary condition in our problem. At first, we focus on the limit $\sigma \to 0$, namely the system of identical oscillators in presence of Gaussian white noise, in which the phase synchronization behaviour on finite-size lattice as well as the possibility of phase transition when the lattice-size tends to infinite, to the best of our knowledge, is still unknown and will thus be addressed in this work. The main idea of our work is to study synchronization in systems of finite-size numerically, and thus we comment on phase transition in such system extending our results of finite-size to thermodynamic limit, if the limit exists, using finite-size scaling analysis. 

Secondly, for completeness of the problem, we revisit the limit $g \to 0$ i.e. the system with quenched noise briefly from a different perspective. We look only at the global synchronization behaviour instead of looking at the local structures, if any, formed in the phase-field of the oscillators, and recover the results for finite system-sizes by performing simple numerical experiments and moreover, arrive at the same conclusion which is already established in the thermodynamic limit. 
We restrict our discussions only to these two limiting cases. The dynamics in presence of both annealed and quenched noise also yields some non-trivial results, which will be reported somewhere else.

From now on, we use the term 'reduced annealed noise-strength' ($g$) and temperature \footnote{We can always write the two-point correlation for Gaussian white noise as $\langle \eta_{i}(t)\eta_{j}({t'}) \rangle = 2T \delta_{ij} \delta(t - {t}^{\prime})$, where T plays the role of temperature. Thus, according to our definition, $g=\sqrt{2T/K}$.} interchangeably and so do for 'reduced quenched noise-strength' ($\sigma$) with 'width' of natural frequency distribution. All the numerical results are obtained simulating the governing dynamics (Equation \ref{govn_eqn_4}) with periodic boundary conditions and the initial distribution of the oscillator phases are chosen to be zero. The steady state properties are calculated by taking time-average over a sufficient time window in the stationary state for each noise realization and finally averaged over 100 such independent realizations.

We study the system the same way as is done in globally coupled system, by defining a (complex valued) macroscopic quantity called "Order parameter"
\begin{equation}
    \rho e^{{\rm i} \psi} = \frac{1}{N} \bigg \langle \sum_{j=1}^{N} e^{{\rm i} \theta_{j}} \bigg \rangle
\label{order_para_def}
\end{equation}
where $\langle \cdot \rangle$ represents averaging over noise realization, $\rho$ quantifies the degree of synchronization and $\psi$ is the average phase of synchronized oscillators. 
\clearpage

\section{Oscillators with annealed noise: Linear theory}
\label{PT:annealed noise}
In this section, we explore the synchronization phenomena in the bare model on finite lattice in presence of annealed noise. We would first study the linearized version of the model, which is analytically tractable, to get some insight about the synchronization in the system. 
In the weak noise-strength regime, we linearize the system by assuming that the phase difference between each pair of neighbouring oscillators is very small ($\Delta \theta \ll 1$). In this approximation, we make the variable $\theta$ unconstrained by extending its range from $(-\pi,\pi)$ to $(-\infty,\infty)$.

Eq. \ref{govn_eqn_4} takes the form (in the limit $\sigma \to 0$)
\begin{eqnarray}
\frac{{\rm d} \theta_{i}}{\rm dt} = \sum_{j \in nn_{i}} (\theta_{j}- \theta_{i}) + g \xi_{i}(t).
\label{govn_eqn_5}
\end{eqnarray}

We derive the single oscillator stationary-state probability distribution in the linear regime using mean-field technique. We replace the local spatial interactions with the global coupling through an average field.
If $W(\{\theta\},t)$ be the multivariate probability density distribution, the Fokker-Planck equation for $W(\{\theta\},t)$ can be written as
\begin{eqnarray}
\frac{\partial W(\{\theta\},t)}{\partial t} = - \sum_{i} \frac{\partial}{\partial \theta_{i}}\left[ \sum_{j \in nn_{i}} (\theta_{j} - \theta_{i})  \right] W(\{\theta\},t) + g^{2} \sum_{i} \frac{\partial^{2}}{\partial \theta^{2}_{i}}  W(\{\theta\},t).
\end{eqnarray}
The probability distribution for a single oscillator can thus be obtained by integrating out the rest of the oscillators as follows:
\begin{eqnarray}
W(\theta_{i},t) = \int \prod_{j \neq i} d\theta_{j} W(\{\theta\},t).
\end{eqnarray}
So the Fokker-Planck equation for single site probability distribution can be written as
\begin{eqnarray}
\frac{\partial W(\theta,t)}{\partial t} = -  \frac{\partial}{\partial \theta}\left[ - D(\theta - E(\theta,t))  \right] W(\theta,t) + g^{2}  \frac{\partial^{2}}{\partial \theta^{2}}  W(\theta,t). \label{single_site_prob_identical_osc}
\end{eqnarray}
where the subscript $i$ has been dropped for notational convenience, $D$($=2d$, for spatial dimension $d$ ) represents the number of nearest neighbours of each lattice site on the lattice and $ E(\theta,t)$ is defined as
\begin{eqnarray}
E(\theta,t) = \int d \theta'\theta' W(\theta'|\theta, t).
\end{eqnarray}
where $W(\theta'|\theta, t)$ is a conditional probability. The steady state probability distribution is obtained by solving Equation \ref{single_site_prob_identical_osc}:\cite{risken2012fokker,garcia2012noise}
\begin{eqnarray}
W_{\rm st}(\theta) = C \exp \left[ \frac{1}{g^{2}} \int_{0}^{\theta} d\psi ( - D[\psi - E_{\rm st}(\psi)]) \right]
\end{eqnarray}
This solution is obtained by making the probability current zero.
Now, we assume that the steady state conditional average $E_{\rm st}(\psi)$ is uniform and thus $E_{\rm st}(\psi) = E_{\rm st} = \langle \theta \rangle$. $C$ is the normalization constant to be fixed by the condition that $W_{\rm s}(\theta + 2 \pi) = W_{\rm st}(\theta)$.
\begin{eqnarray}
 W_{\rm st}(\theta) = C \exp \left[ - \frac{1}{g^{2}} ( \frac{D}{2} [\theta^{2} - 2 \theta E_{\rm st}]) \right]
& = C' \exp \left[ - \frac{D}{2g^{2}} (\theta - \theta_{\rm m} )^{2} \right]
\end{eqnarray}
where $\theta_{\rm m} = E_{\rm st} $ and $C'= C \exp(\frac{D \theta_{\rm m}^{2}}{2g^{2}})$.\\
Thus, a single site equilibrium stationary-state probability distribution is Gaussian centered at $\theta_{\rm m}$ with variance $g^{2}/D$. Of course, this distribution is true only for weak noise-strength (low temperature) in systems of finite size.



\begin{figure}[]
\hspace{-1.0 cm}
\includegraphics[scale=0.6]{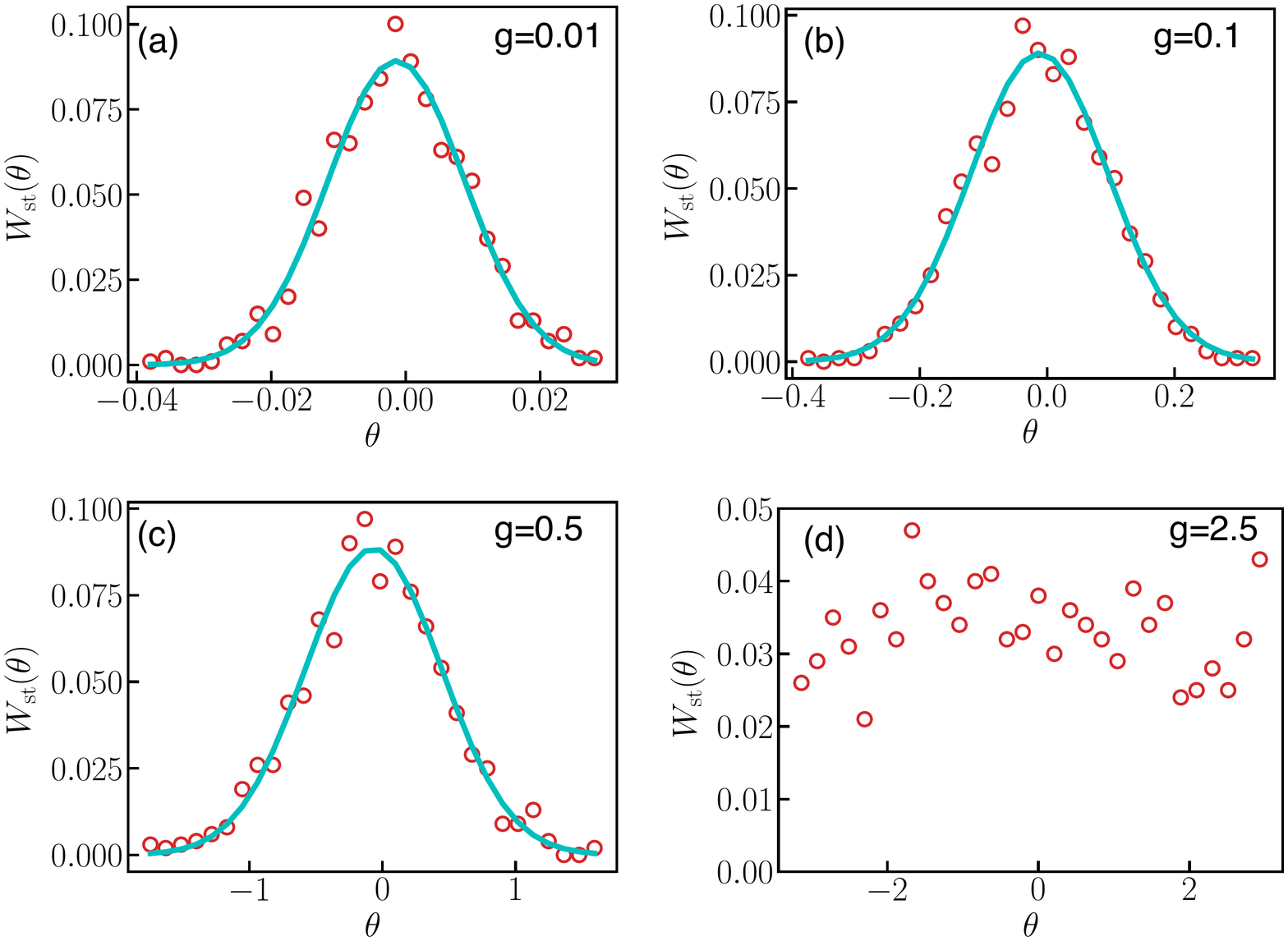}
\caption{(Color online) Single oscillator stationary-state probability distribution for a lattice of size $N= 50\times 50$ is shown for different noise-strengths $g=0.01,0.1,0.5$ and $2.5$ in (a), (b), (c) and (d) respectively. The initial phases are chosen to be zero and the distribution is obtained after averaging over $1000$ independent runs. The distribution is Gaussian in linear regime ((a),(b) and (c)) whereas it tends to a uniform one at high noise-strength (d) where linear approximation is not valid.}
\label{fig:single_oscillator_prob_distribution}
\end{figure}

To verify the above results, the probability distribution of a single oscillator on the finite-size lattice in the stationary state is calculated at different noise-strengths. These are shown in Figure \ref{fig:single_oscillator_prob_distribution}. At very small noise-strengths where the linear approximation is valid, they appear to be gaussian (Figure \ref{fig:single_oscillator_prob_distribution} (a), (b) and (c)). The width of the distribution increases with the noise-strength linearly in this region and finally the distribution becomes uniform at very high noise-strength (Figure \ref{fig:single_oscillator_prob_distribution}(d)).
	
We note that, the linear theory predicts logarithmic divergence of the mean phase-variance in thermodynamic limit at any noise-strength (Appendix \ref{appendix_EW}). So, in the framework of linear theory, we can not comment on existence of any kind of phase transition in this system in thermodynamic limit. In the strong noise-strength limit, the nearest neighbour phase difference may become unbounded even in finite dimension and the oscillator phases become completely random. So the system of finite size shows a crossover from a state where $\rho =1$ to a state where $\rho \sim N^{-1/2}$ implying complete disordered state. To understand the complete picture, the full non-linear system is taken into consideration and is investigated numerically in the next section.


\section{Oscillators with annealed noise: Phase-synchronization transition}
\label{PT:KT}
In this section, we study numerically the phase-synchronization behaviour of the full non-linear system by studying order parameter and other statistical quantities with temperature in finite systems of various sizes. The behaviour of these statistical quantities with the system-sizes shows 
the signature of true phase transition in the thermodynamic limit. 
\subsection{Order parameter}
\begin{figure}[]
\begin{subfigure}{.49\textwidth}
\hspace{-1.0 cm}
\includegraphics[scale=0.7]{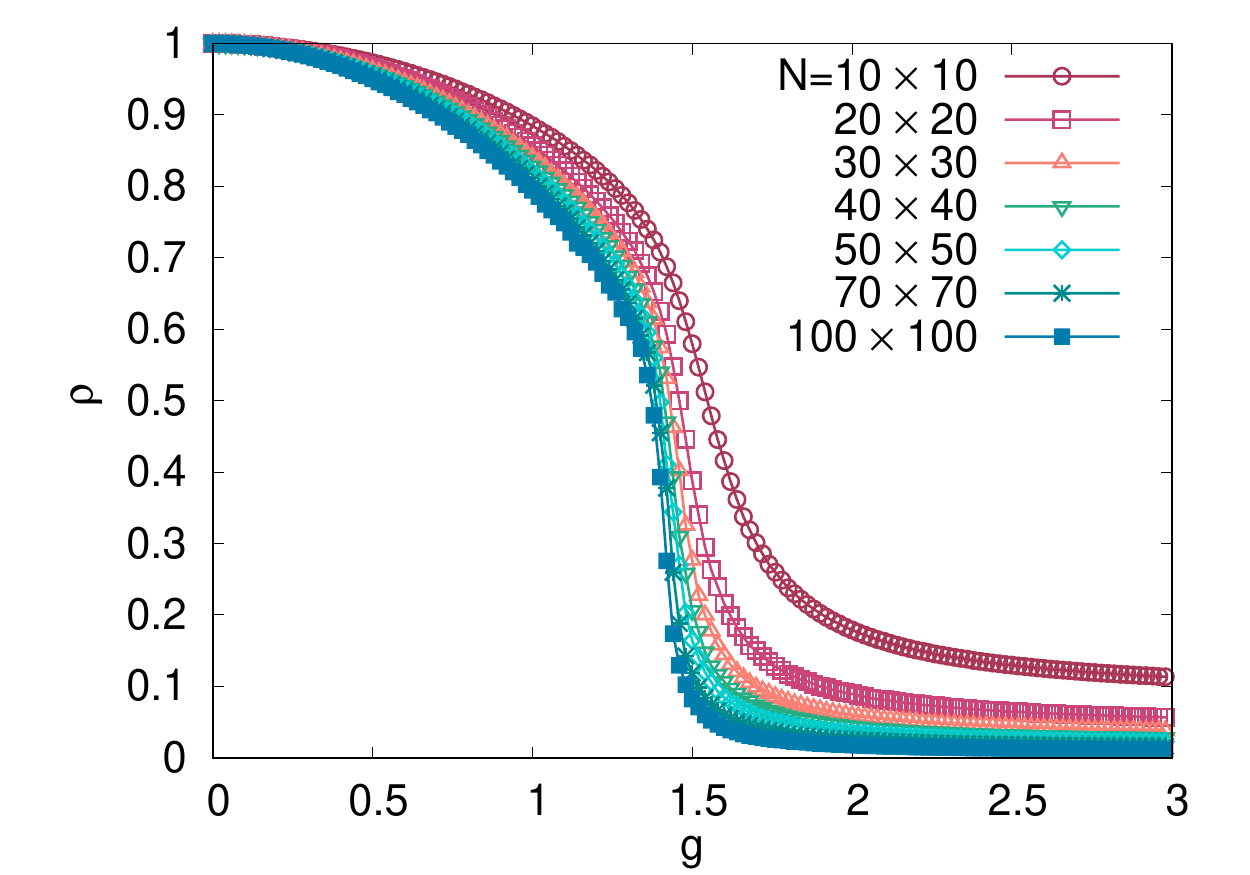}
\caption{}
\end{subfigure}
\begin{subfigure}{.49\textwidth}
\hspace*{-0.5 cm}
\includegraphics[scale=0.7]{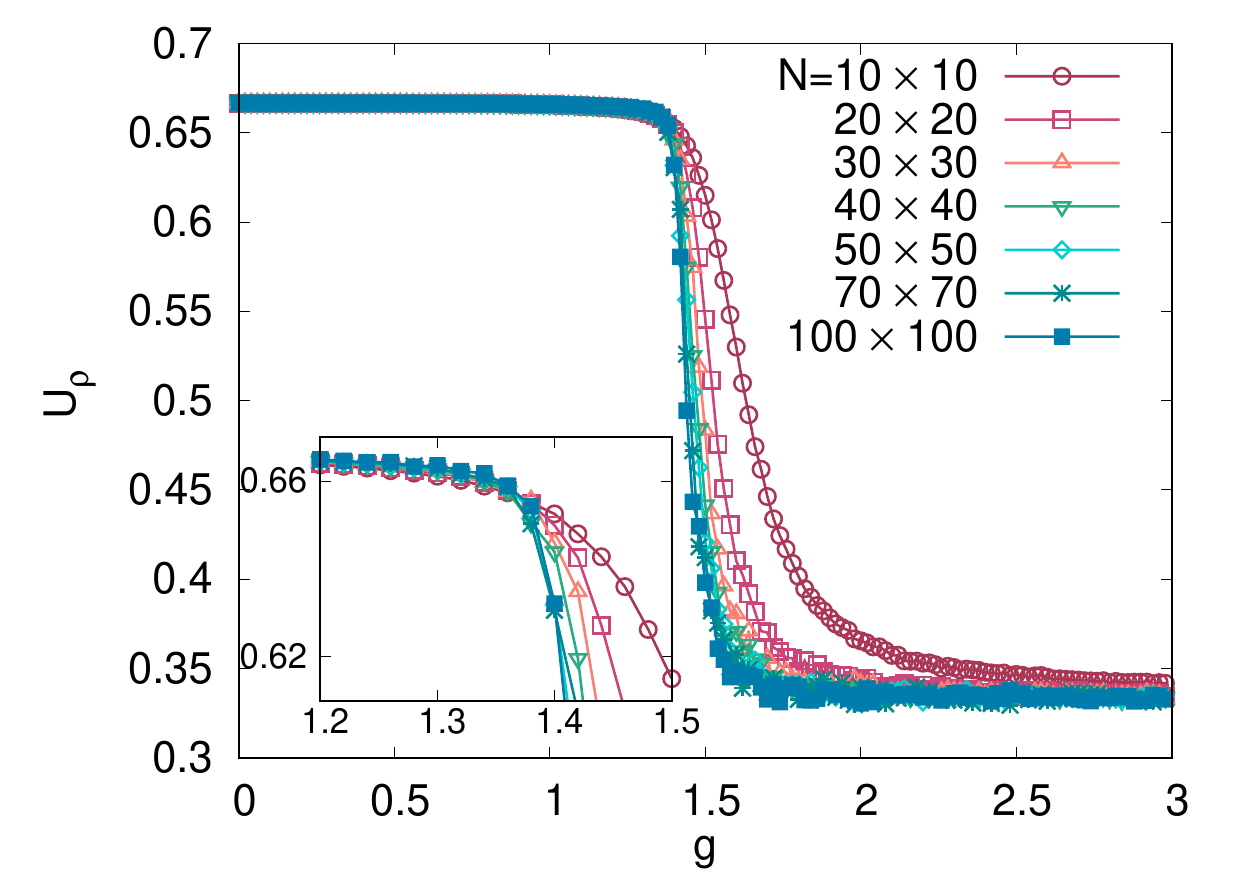}
\caption{}
\end{subfigure}
\caption{(Color online) Variation of order parameter ($\rho$) and Binder cumulant ($U_{\rho}$) with the noise-strength ($g$) for various system sizes ($N=L \times L$) is shown in (a) and (b) respectively. The order parameter seems to exhibit a continuous transition from synchronization to de-synchronization with the noise-strength indicating a phase transition in the thermodynamic limit. Inset: The curves $U_{\rho}$ for various $L$ seem to stay collapsed upto $g=g_c$ and then separate. This shows the existence of critical phase (with power-law decay), which is a signature of $KT$ transition in the thermodynamic limit.}
\label{fig:gop_Binder_cumulant_variation_with_g_different_syst_sizes}
\end{figure}

Here, we investigate the variation of phase order parameter, as defined in Equation \ref{order_para_def}, of the full non-linear system as noise-strength varies. Figure \ref{fig:gop_Binder_cumulant_variation_with_g_different_syst_sizes}(a) shows the behaviour of global phase order parameter ($\rho$) with noise-strength ($g$) for various system sizes ($N=L \times L$).

According to equilibrium critical phenomena, as the parameter changes, the phase order parameter ($\rho$) changes continuously from a non-zero value to zero in the thermodynamic limit. In this limit, we expect $\rho$ to be zero in super-critical region ($g > g_{\rm c}$). Clearly, for large but finite population of oscillators ($N$), $\rho \sim L^{-{\alpha}}$ at the transition point ($\epsilon = 0$) whereas in de-synchronized regime $\rho \sim L^{-d/2}$. So the decay exponent $\alpha$ can be estimated from the system size dependency of the order parameter at criticality.

\begin{figure}[]
\includegraphics[scale=0.35]{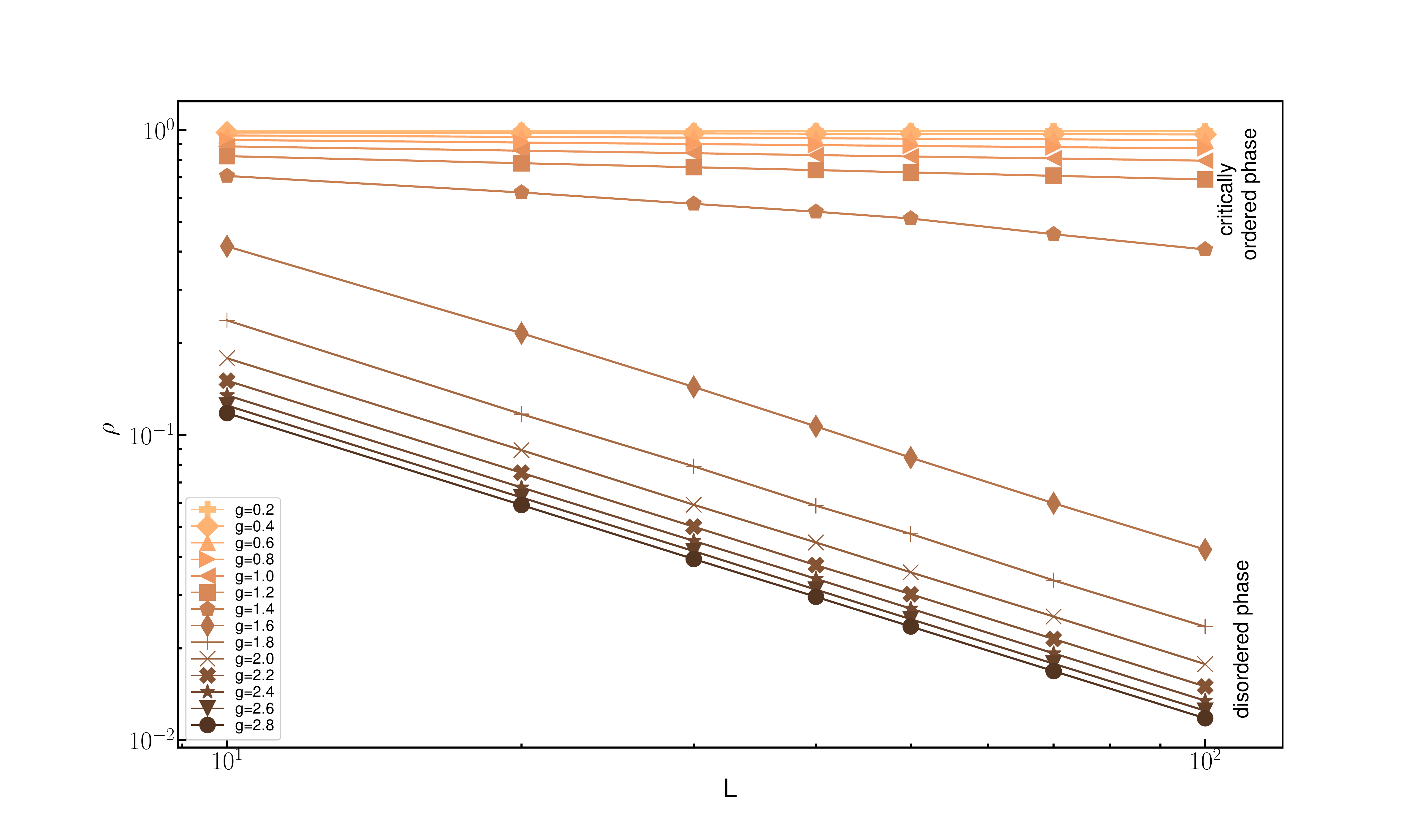}
\caption{(Color online) Scaling of $\rho$ with the system-size $L$, on a logscale, at various $g$-values is shown. In the disordered phase, the exponent is a constant $d/2(=1$, d being spatial dimension), as expected; whereas in the critically ordered phase, the exponent varies with temperature. This suggests the absence of any macroscopic ordering in the system at all temperatures in the thermodynamic limit. }
\label{fig:order_para_KT}
\end{figure}

Interestingly, we observe an unusual behaviour in the steady-state value of $\rho$. It scales with the system size ($L$) with an exponent 1 ($= d/2$) in the de-synchronized regime as expected. But, the scaling with $L$ continues in the synchronized regime also, with temperature dependent $\alpha$ suggesting absence of any macroscopic ordering in the system in the thermodynamic limit. The Table \ref{Table:ord_para_fluc_exponents} records the $\alpha$-values at different temperatures. Lower the temperature, slower is the decay of $\rho$. Overall, the system remains in a critically ordered phase in the region $g \leq g_{c}$. Figure \ref{fig:order_para_KT} shows how $\rho$ scales with $L$ at different $g$-values on a logscale. 
It shows that the scaling holds in both the regions.
We note that the value of the exponent $\alpha$ at very low temperatures is surprisingly very small. One may argue that $\rho$ can then be assumed to get saturated at some finite value. But we emphasize the importance of this exponent. The $\alpha$ values, being small, indicates a slower decay but can not be neglected, and we believe this is due to the strong finite-size effect. 

Thus, the systems of finite-size although exhibit a continuous transition from non-zero order parameter value to very small ($\rho \sim L^{-d/2}$) value, the scaling behaviour with system sizes suggests existence of critically ordered phase for a range of temperatures $0 < g \leq g_c$. Due to the difficulties in predicting the value of $g_c$ from the order parameter variation only, other quantities are also studied which are described in the following sections. 
	
\subsection{Binder Cumulant}
To understand the nature of the transition, another useful diagnostic tool, the fourth order Binder cumulant is measured which is defined by
\begin{eqnarray}
U_{\rho} = 1- \left[\frac{\langle \rho^4 \rangle}{3\langle \rho^2 \rangle^2}\right]
\end{eqnarray}
where $\langle \cdot \rangle$ and $[\cdot]$ represent the time average in the stationary state and sample averages, respectively. Here sample average means taking average over different noise realizations. For large system-size ($N$), in the synchronized regime (weak noise-strength limit), $U_{\rho} \to 2/3$ whereas in the de-synchronized regime, $U_{\rho} \to 1/3$. In both the region, the correlation length $\xi \ll L$ and thus $U_{\rho}$ for various system-sizes remains close to those fixed point values. So, the critical parameter value at which $\xi \to \infty$ can be identified by looking for the common intersection point of the curves for $U_{\rho}$ for various system-sizes.

Figure \ref{fig:gop_Binder_cumulant_variation_with_g_different_syst_sizes}(b) shows the variation of $U_{\rho}$ with noise-strength for various values of $N$. They seem not to intersect through a common point but to collapse and remain so for the noise-strength in a range from zero to a certain critical one ($g_{c} \approx 1.34$). This indicates the existence of critically ordered phases in this range ($g \leq g_{c}$) and also implies the diverging correlation length in this region. Beyond this region ($g > g_{c}$), the curves separate suggesting the onset of disordering and at very high $g$ value, they again take another fixed point value corresponding to complete disordered phases. 
Thus $U_{\rho}$ yields an estimation of $g_{c}$ as well as the nature of the ordered phase.

Now, to estimate $g_{c}$ and critical exponents associated with the transition accurately, we perform finite-size scaling of dynamic fluctuation which will be discussed in the following section.

\subsection{Dynamic Fluctuations}


\begin{figure}[]
\begin{subfigure}{.49\textwidth}
\vspace*{+0.5 cm}
\hspace{-1.5 cm}
\includegraphics[scale=0.8]{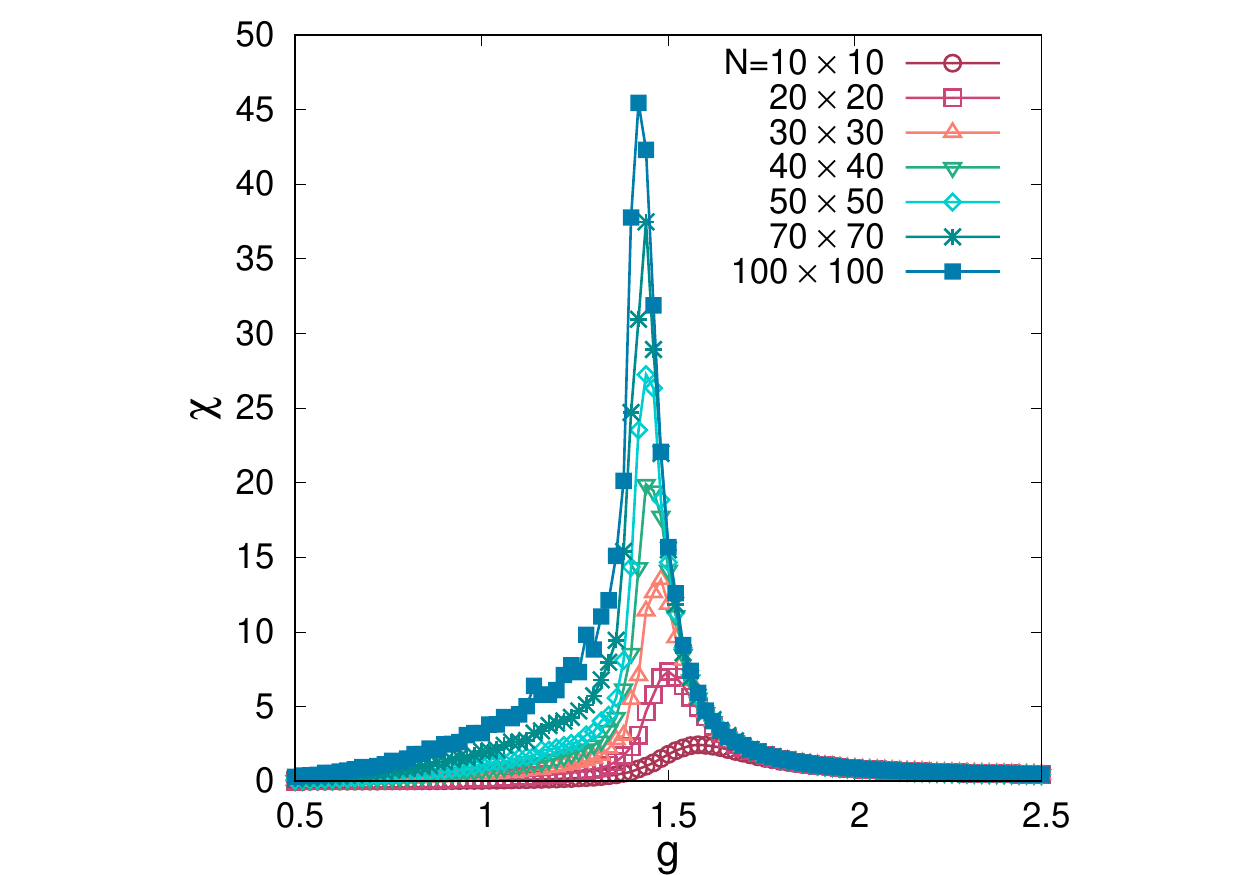}
\caption{}
\end{subfigure}
\begin{subfigure}{.48\textwidth}
\hspace{-0.5 cm}
\includegraphics[width=9 cm, height=7.7cm]{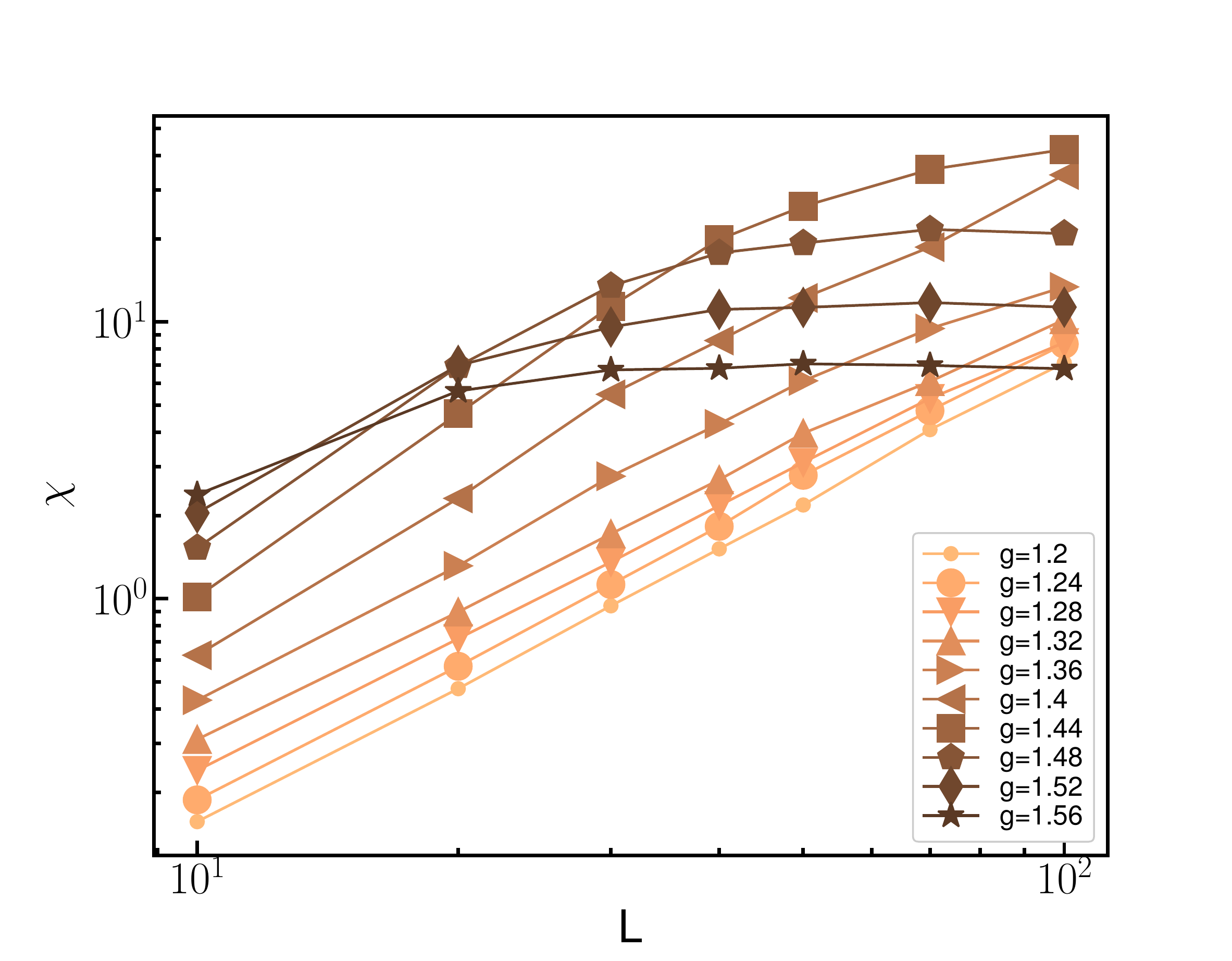}
\caption{}
\end{subfigure}
\caption{(Color online) Variation of dynamic fluctuation ($\chi$) with the noise-strength ($g$) for various system sizes ($N=L \times L$) is shown. The height of the peaks increases for larger system sizes indicating a faster divergence at high $g$-values in the thermodynamic limit. (b) Scaling of $\chi$ with the system-size $L$, on a logscale, at various $g$-values is shown. This scaling holds upto $g = g_c$, suggesting the existence of critically ordered phase in this region. Beyond $g_c$, the fluctuations becomes finite.}
\label{fig:dyn_fluc_variation_with_g_different_syst_sizes_scaling_with_L_near_g_c}
\end{figure}

 To estimate the $g_c$, we measured the fluctuation of the order parameter defined by
\begin{eqnarray}
\chi_{\rho} (g, L) = L^{d}[\langle \rho^2 \rangle - \langle \rho \rangle^2]
\end{eqnarray}
where $\langle \cdot \rangle$ and $[\cdot]$ represent the time average in the stationary state and sample averages, respectively. 
From now on, we will drop the subscript ${\rho}$ in $\chi_{\rho}$. This quantity is equivalent to susceptibility in statistical system.

Variation of dynamic fluctuation ($\chi$) with the noise-strength ($g$) for various system sizes ($N=L \times L$) is shown in Figure \ref{fig:dyn_fluc_variation_with_g_different_syst_sizes_scaling_with_L_near_g_c}(a). The height of the peaks increases for larger system sizes and the peak-positions shift towards lower values of $g$. The maxima of the fluctuations also scale with linear system-size, indicating diverging correlation length at some critical $g$ in the thermodynamic limit. So far, this resembles the conventional phase transition of second order. But, the story does not end here. In fact, the existence of critically ordered phase for a certain range of $g$ also gets reflected in the behaviour of $\chi$ with temperature. Figure \ref{fig:dyn_fluc_variation_with_g_different_syst_sizes_scaling_with_L_near_g_c}(b) shows the variation of $\chi$ with $L$ at various $g$-values, on a logscale. For $g < g_c$, $\chi$ scales with $L$ with a temperature dependent exponent, beyond which the scaling does not hold. Thus, in the thermodynamic limit, the susceptibility is infinite at all temperatures below $g_c$ (in the critically ordered phase). It is also evident that this scaling with $L$ fails as $g$ exceeds $g_c$ and finally, at very high temperatures, the fluctuation becomes saturated at some finite value $\mathcal{O}(1)$.


 
 \begin{table}[]
    \centering
    \begin{tabular}{| P{0.75 cm} | P{1.25 cm}| P{1.25 cm} || P{0.75 cm} | P{1.25 cm} | P{1.25 cm} |}
    \hline
    $g$ & $\alpha$ & $\gamma$ & $g$ & $\alpha$ & $\gamma$\\
    \hline
    0.2 & 0.0016(0) & 1.98(0) & 1.2 & 0.0763(2) & 1.67(3)\\
    0.4 & 0.0065(1) & 1.84(2) & 1.6 & 0.979(9) & -\\
    0.6 & 0.0150(1) & 1.72(3) & 1.8 & 1.008(4) & -\\
    0.8 & 0.0276(3) & 1.71(3) & 2.0 & 1.004(1) & -\\
    1.0 & 0.0469(2) & 1.73(2) & 2.2 & 1.003(0) & -\\
    \hline
    \end{tabular}
    \caption{Temperature dependence of the exponents $\alpha$ and $\beta$ is listed. The estimated error is shown in parentheses.}
    \label{Table:ord_para_fluc_exponents}
\end{table}
 
 So far the behaviour of the order parameter, Binder cumulant and dynamic fluctuations observed in finite systems suggests the existence of critically ordered phase in the region $0 < g \leq g_c$ and disordered phase in the region $g > g_c$ in the thermodynamic limit. For the range of temperatures over which the critically ordered phases exist, critical indices vary continuously with temperature. This phase transition from critically orderd phase to disordered phase is analogous to the Kosterlitz-Thouless ($KT$) transition as observed in $2D$ $XY$ model \cite{kosterlitz1973ordering,kosterlitz1974critical}. Actually, there is a relationship between the Kuramoto model under study and the classical $XY$ model of statistical system, which is established in Appendix \ref{appendix_KM_XY_equivalence}. The bare model with annealed noise on $2D$ lattice is equivalent to overdamped dynamics of the classical $2D$ $XY$ model with nearest neighbour interaction in contact with a heat reservoir and the stationary state dynamics is thus governed by equilibrium statistical mechanics. The $KT$ transition and the critical behaviour near the transition in the $2D$ $XY$ model in the thermodynamic limit are briefly mentioned in Appendix \ref{appendix_KT}. 
 
 Now, we perform finite-size scaling to estimate the transition temperature in our system in the thermodynamic limit. We assume that, the correlation behaviour would also be similar to that of $2D$ $XY$ model (Equation \ref{Eq:chi_xy_KT}). For a system of finite size, the correlation length, at criticality, can be assumed to be of the order of the linear size of the system $\xi \sim L $. 
Thus, based on our assumptions, the following relation holds:
\begin{eqnarray}
|g_c(\infty) - g_c (L)| \propto (\log L)^{-1/\nu}
\end{eqnarray}
where $g_c(\infty)$ and $g_c{(L)}$ are the critical noise-strengths in the thermodynamic limit and in finite system of size $N(=L^{d})$, respectively. Here, $\nu$, which is 0.5 for $2D$ $XY$ model, is the exponent which determines how fast $\chi$ would diverge. The quantity $g_c (L)$ can be calculated from the position of maxima of dynamic fluctuations.

Figure \ref{fig:finding_g_c_spatial_corr_KT}(a) shows the system-size dependency of the critical noise-strength. The best fitted curve through the datapoints yields the exponent $\nu$ and critical noise-strength in the thermodynamic limit 
\begin{eqnarray}
1/\nu = 1.58(14) \qquad\text{and}\qquad g_c (\infty) = 1.338(14),
\end{eqnarray}
which, in terms of temperature, turn out to be,
\begin{equation}
\nu \approx 0.63 \qquad \text{and} \qquad T_c \approx 0.895 \qquad{\text{in units of $K/k_B$}},
\end{equation}
The exponent $\nu$ deviates from the theoretical $KT$ prediction but, the critical temperature $T_c$, although deviates from actual prediction, is in agreement with the transition temperature, given by ${k_B T_c} /K \approx 0.89$, obtained via Monte Carlo simulation on the $2D$ $XY$ model \cite{tobochnik1979monte,fernandez1986critical}.


\subsection{Two-point Correlation}
As already discussed in the previous section, in the region $g \leq g_{c}$, the system remains in critically ordered phase i.e. $\xi$ is infinite in the thermodynamic limit which in turn implies the power-law behaviour of correlation in this phase. As a final verification, we calculate equal-time connected correlation function (spatial correlation), defined as
\begin{equation}
C(r,0)= \langle \theta(0).\theta(r) \rangle_{c} = \Re{\langle \exp (\rm i (\theta(0)-\theta(r)) \rangle} - \rho^{2}
\end{equation}
where $\langle \cdot \rangle$ represents averaging over oscillators. It is computed in the following way: first circular bins are formed around an oscillator in a particular steady state configuration, ${\langle \exp ({\rm i} (\theta(0)-\theta(r)) \rangle}$ is calculated for each bin and then the same process is repeated for each oscillator in that configuration and finally the averaging is done over all the oscillators. This whole process is further averaged over sufficient number of configurations for each run and finally averaged over 100 such independent runs.

\begin{figure}[]
\begin{subfigure}{.49\textwidth}
\hspace{-1.5 cm}
\includegraphics[scale=0.8]{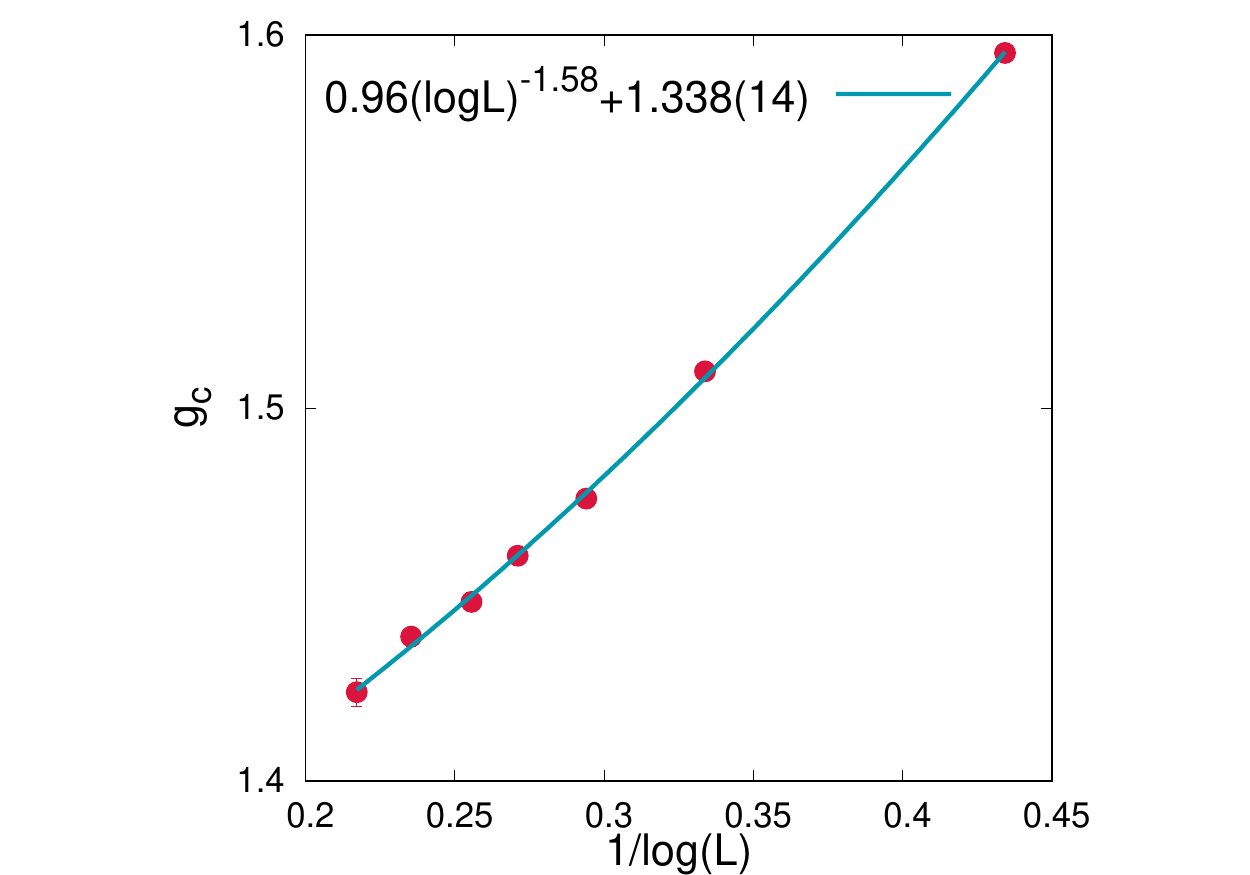}
\caption{}
\end{subfigure}
\begin{subfigure}{.49\textwidth}
\hspace{-1.5 cm}
\includegraphics[scale=0.8]{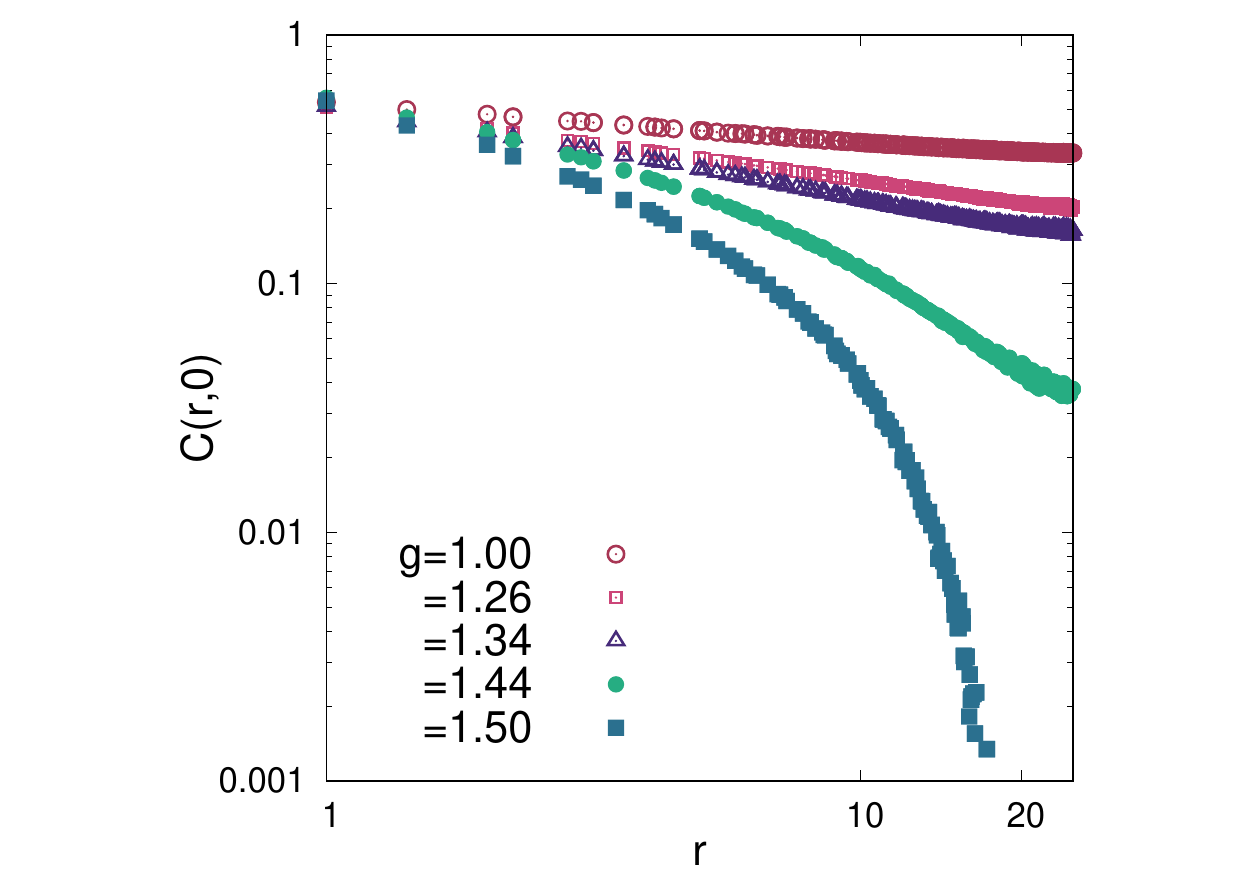}
\caption{}
\end{subfigure}
\caption{(Color online) (a) Variation of critical noise-strength ($g_c$) with  linear system size ($L$) is shown. The data-points are plotted with the error-bars. In some cases, the error-bar is smaller than the size of the data-point. The $y$-intercept of the fitted line gives the value of critical noise-strength in the thermodynamic limit. (b) Behaviour of Spatial correlation, on a log-scale, at different temperature is shown for a system of size $N=50\times50$.}
\label{fig:finding_g_c_spatial_corr_KT}
\end{figure}

Figure \ref{fig:finding_g_c_spatial_corr_KT}(b) shows the behaviour of $C(r,0)$, on a log-scale, at different temperatures for a particular system size $N=50 \times 50$. The distance $r$ over which correlations are observed is restricted to $r \approx L/2$ due to finite-size effects. For $g \leq g_c$, there is an algebraic decay of the correlation. The power-law exponent ($\eta$) is calculated at different temperatures and is listed in Table \ref{Table:eta_temp}.
However, in the de-synchronized region $g >g_c$, the correlation falls off exponentially fast. The critical exponent $\eta$ at $g \approx g_c$ is found to be $\eta = 0.39$ which differs from the value 0.25 predicted from $KT$ theory. 

\begin{table}[]
    \centering
    \begin{tabular}{| P{1.5 cm} | P{3 cm}|| P{1.5 cm} | P{3 cm}|}
    \hline
    $g$ & $\eta$ & $g$ & $\eta$ \\
    \hline
    0.5 & 0.045(0) & 1.26 & 0.298(0)\\
    1.0 & 0.147(1) & 1.34 & 0.387(1)\\
    \hline
    \end{tabular}
    \caption{Temperature dependence of the exponent $\eta$ for a system of size $N=50 \times 50$ is listed. The estimated error is shown in parentheses.}
    \label{Table:eta_temp}
\end{table}

We observe sufficient deviation in the value of the critical exponents from the KT predictions. In fact, the estimation of the exponents numerically for the $2D$ $XY$ model itself is a difficult task and requires computation on very large system-size with very minute observations; otherwise, a little error introduced in the fitting may lead to a large deviation and even may change the order of the transition \cite{gupta1988phase}. In our case, we believe this deviation may arise due to the small system-size taken into consideration and thus may suffer from large finite-size effects. The study on a very large system-size being computationally expensive, we restrict ourselves to a smaller one. We emphasize that more than the exact value of the exponents $\nu$ and $\eta$, we are interested in unveiling the nature of the phase transition in our system via numerical investigations on finite system-sizes.


        To sum up, the bare Kuramoto oscillators on a $2D$ square lattice in presence of annealed noise exhibits a true phase transition in the thermodynamic limit, which is $KT$-type, where the system makes a transition from a critically ordered phase to a disordered one at the critical noise-strength $g_c \approx 1.34$, or equivalently, critical temperature $T_c \approx 0.90$, in agreement with the $KT$ prediction based on Monte Carlo simulation. In terms of correlations, the transition is from a phase with quasi-long range order, characterized by an algebraic decay of correlations to a disorder where the decay is exponential. The critical exponents are found to be 
        \begin{equation*}
            \nu = 0.63 \qquad \text{and} \qquad \eta = 0.39 \qquad\text{(approx.)}.
        \end{equation*}
        
\clearpage

\section{Oscillators with quenched noise: Synchronization crossover}
\label{PT:quenched noise}
In the present section, we review the synchronization phenomena in the bare model in presence of quenched noise, the dynamics of which is given by \begin{equation}
\frac{{\rm d} \theta_{i}}{\rm dt} =  \sigma \omega_{i} + \sum_{j \in nn_{i}}  \sin(\theta_{j}- \theta_{i})
\label{govn_eqn_quenched_noise}
\end{equation}
To characterize the synchronization behaviour, we measure the same statistical quantities as before as a function of quenched noise-strength (or equivalently, width of distribution of the natural frequencies, $\sigma$) which is described in the next section. 
\subsection{Order Parameter, Binder Cumulant and Dynamic Fluctuations}

\begin{figure}[]
\begin{subfigure}{.49\textwidth}
\hspace{-1.5 cm}
\includegraphics[scale=0.8]{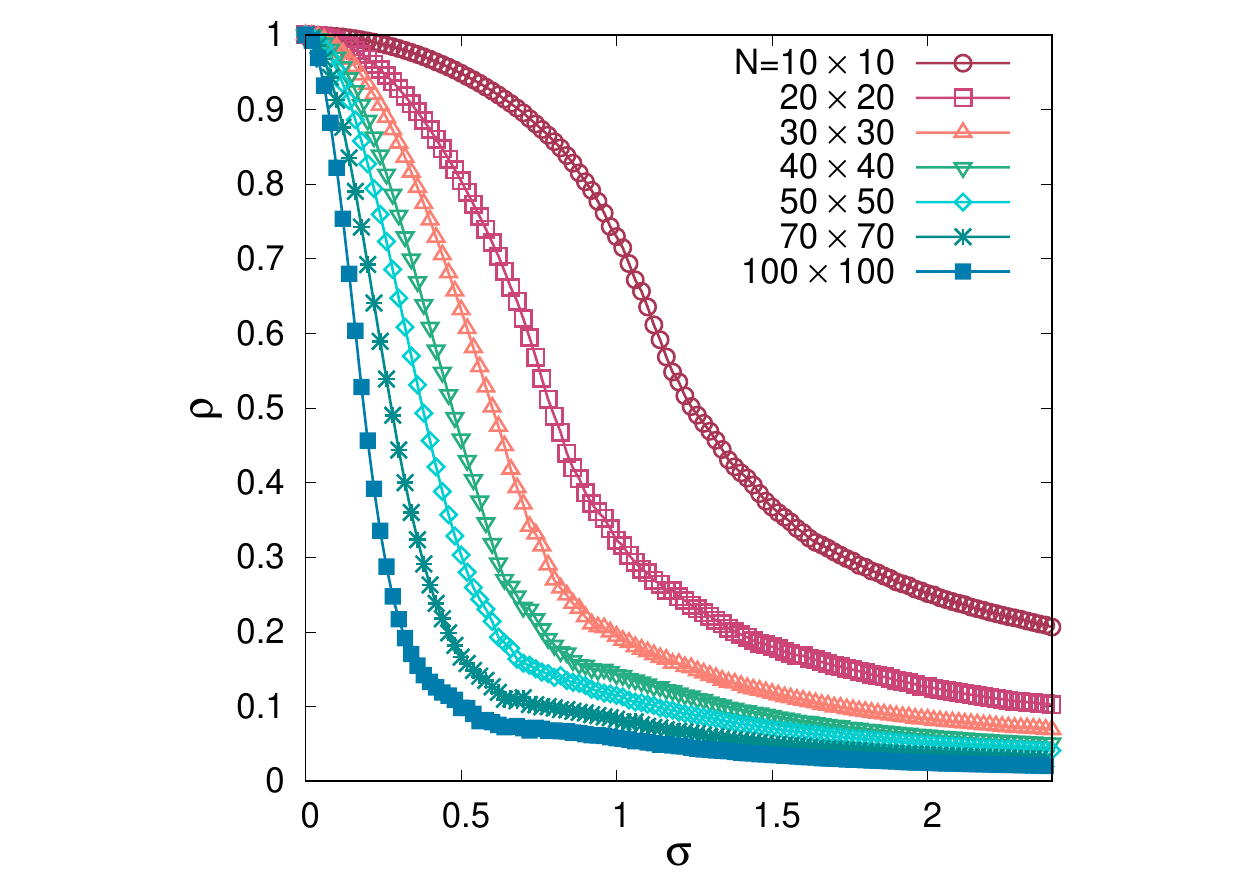}
\caption{}
\end{subfigure}
\begin{subfigure}{.49\textwidth}
\hspace*{-1.5 cm}
\includegraphics[scale=0.8]{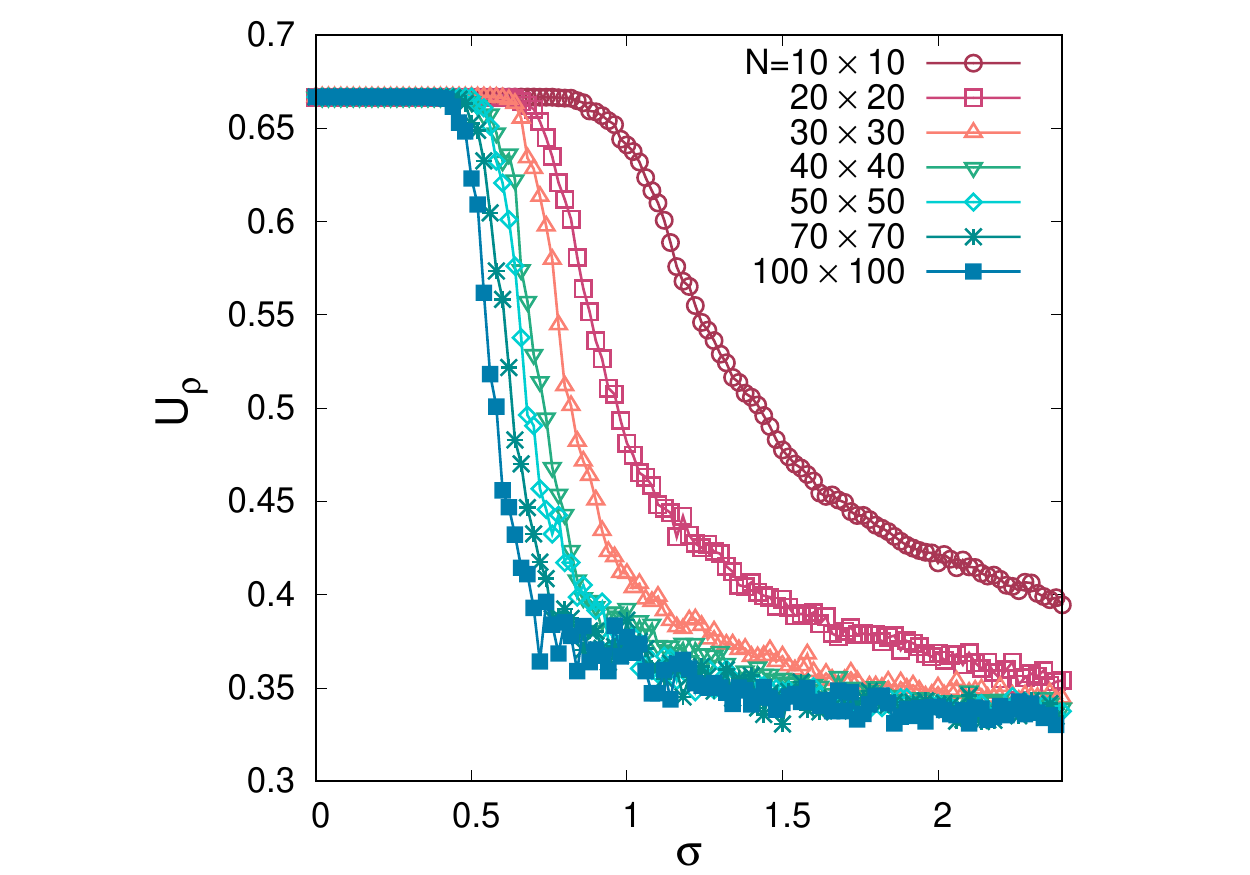}
\caption{}
\end{subfigure}
\caption{(Color online) (a) variation of global order parameter ($\rho$) with quenched noise-strength ($\sigma$) is shown for various system-sizes ($N=L \times L$). (b) Variation of Binder cumulant ($U_{\rho}$) with $\sigma$ for various $N(=L \times L)$ is shown. The curves for $U_{\rho}$ do not have a common intersection point which indicates the absence of phase transition in the thermodynamic limit.}
\label{fig:order_para_Binder_cumulant_vs_sigma_quenched_noise}
\end{figure}

The order parameter ($\rho$) variation with $\sigma$ shows that the system exhibits a crossover from synchronization to desynchronization as $\sigma$ varies. For a particular system size($N=L \times L$), $\rho$ decreases continuously with the increase of $\sigma$ and finally the system becomes completely desynchronized at high $\sigma$-values. Figure \ref{fig:order_para_Binder_cumulant_vs_sigma_quenched_noise}(a) shows the behaviour of $\rho$ with $\sigma$ for various values of $L$, showing a crossover from synchronization to de-synchronization for finite $L$. The system of larger size gets desynchronized for lower values of $\sigma$. 

To understand, whether this crossover exists in the thermodynamic limit, we look at the behaviour of Binder's fourth cumulant ($U_{\rho}$) with $\sigma$ for different $L$ values and is shown in the Figure \ref{fig:order_para_Binder_cumulant_vs_sigma_quenched_noise}(b). The crossover from synchronization to desynchronization in finite $L$ is evident, but the curves of $U_{\rho}$ for various $L$ are distict and do not intersect through a common point. Clearly, this is a strong evidence in support of the absence of any phase transition in the thermodynamic limit. 

As a final verification, we study the dynamic fluctuations as a function of $\sigma$, and calculate the crossover noise-strength $\sigma_c (L)$ from its maxima for various system-sizes ($N=L \times  L$). Figure \ref{fig:chi_vs_sigma_and_sigma_c_with_L_quenched_noise}(a) shows the behaviour of $\chi$ with $\sigma$ for various $L$, which shows that the maxima of the fluctuations shift towards lower value of $\sigma$ for large $L$-values. The maximum value of the dynamic fluctuation ($\chi_{\rm max}$), as shown in the inset of Figure \ref{fig:chi_vs_sigma_and_sigma_c_with_L_quenched_noise}(a) on a logscale, does not follow any power law scaling with the linear system-size ($L$), which suggests that the correlation length is finite as the system "crosses over" from synchronization to desynchronization. We show the system size dependency of $\sigma_c (L)$ in Figure \ref{fig:chi_vs_sigma_and_sigma_c_with_L_quenched_noise}(b) on a semi-log scale, where the red circles represent reciprocal of $\sigma_{c}(L)$ values for different $L$. The best fitted curve suggests that $\sigma_c$ scales with $L$ as  $\sigma_{c} \propto 1/\log(L)$ and thus tends to zero as $L \to \infty$. Thus, although the system of finite $L$ does exhibit a crossover but it does not in the thermodynamic limit. Figure \ref{fig:chi_vs_sigma_and_sigma_c_with_L_quenched_noise}(c) shows the system-size dependency of $\sigma_c$ (circles in red) calculated directly from simulation as described above.

At the oscillator level, the picture is as follows: initially, in absence of any noise, all the oscillators are having the same phase, thus the system is in complete phase synchronized state. As we introduce a small disorder, all the oscillators adjust their phases accordingly and become phase-locked, suggesting a decrease in $\rho$-value. That means, all the oscillators now evolve with the same frequency. This phase is thus called "Entrained phase". As the quenchedness increases, the phases become more random, but remain phase-locked and different features are observed in the phase-field of the oscillators, which is not subject matter of our study. This happens upto certain value of $\sigma$, then on further increasing of $\sigma$, few oscillators go out of synchrony while the rest of them remain phase-locked. These oscillators are called 'runaway' oscillators. As we keep on increasing $\sigma$ further, the number of such runaway oscillators increase and finally there emerges a completely desynchronized phase (so called "unentrained phase") at high $\sigma$-values.

From our study of different statistical quantities, we conclude that, this synchronization desynchronization crossover does not exist in the thermodynamic limit. Thus, this system does not show any conventional phase transition; in other words, phase-locking is not possible in such system in the thermodynamic limit which is validated by the earlier works \cite{strogatz1988phase,hong2005collective}.

In the next section, we discuss the linear stability of the stationary solutions and calculate the crossover noise-strength $\sigma_c$ based on the above discussion.

\subsection{Instability of de-synchronized solution (Semi-analytical stability analysis)}
We perform standard linear stability analysis to investigate the stability of the stationary solutions, both the synchronized and the de-synchronized ones. By stationary, we mean that the distribution of the phases of the oscillators, irrespective of whether they are synchronized or desynchronized, remains same in time. We must mention that, in the stationary state of a desynchronized solution, even if the different oscillators have different angular velocities and that too changes in time, the distribution of their phases remains unchanged.
If $\{\theta_0\}$ be one such stationary state solution, we can introduce a small perturbation ($\delta \theta_{0}$) to it. In the linear approximation, the perturbation grows following the equation
\begin{equation}
\delta \dot{ \rm{ \bm\theta}_{0}} ={\bm {J}}(\theta,\omega) \delta{\bm \theta_{0}},
\end{equation}
where $\delta { \bm\theta}_{0}$ is the perturbation vector and ${\bm {J}}(\theta,\omega)$ is the Jacobian matrix evaluated at that point, the form of which is given by
\begin{equation}
{\bm{J}_{ij}} = \frac{\partial}{\partial \theta_{j}} F(\theta_{i},\theta_{j})\bigg|_{\{\theta_0\}},
\end{equation}
where $F(\theta_{i},\theta_{j}) =  \sigma \omega_{i} + \sum_{j \in nn_{i}}  \sin(\theta_{j}- \theta_{i}).$
Explicitly, for our $2D$ model,
\begin{multline}
{\bm J}_{ijlm} =   \bigg[ \cos(\theta_{i+1j}- \theta_{ij}) \delta_{i+1l}\delta_{jm}  +   \cos(\theta_{i-1j}- \theta_{ij}) \delta_{i-1l}\delta_{jm} \\
+  \cos(\theta_{ij+1}- \theta_{ij}) \delta_{il}\delta_{j+1m}  +  \cos(\theta_{ij-1}- \theta_{ij}) \delta_{il}\delta_{j-1m} \\
- \Big (  \cos(\theta_{i+1j}- \theta_{ij}) + \cos(\theta_{i-1j}- \theta_{ij}) \\
+ \cos(\theta_{ij+1}- \theta_{ij}) + \cos(\theta_{ij-1}- \theta_{ij}) \Big ) \delta_{il}\delta_{jm} \bigg] \bigg|_{\{\theta_0\}}.
\end{multline}
Obviously, ${\bm J}_{ijlm} = {\bm J}_{lmij} \forall i,j,l,m.$, i.e. the matrix is symmetric. Thus all the eigenvalues must be real. 
The eigenvalues of this Jacobian determine the local stability of the states. If any of the eigenvalues are positive, this state is then locally unstable. It seems the Jacobian has no explicit dependence on the parameter $\sigma$, so do its eigenvalues. Actually, the effect of $\sigma$ lies in the solutions obtained, at which the the Jacobian is evaluated. The eigenvalues of the Jacobian is evaluated numerically at both the synchronized as well as desynchronized states. There always exists a zero eigenvalue corresponding to the translational symmetry of the system. By translational symmetry, we mean, the system remains invariant under a linear transformation $\{ \theta_i \} \to \{ \theta_i + \alpha\}$. For $\sigma \leq \sigma_c$, all other eigenvalues are negative implying that the synchronized states have all locally stable directions except one, which is neutrally stable. Given an initial condition, the system admits only one stable phase-locked solution corresponding to a particular realization of quenched disorder.

But as $\sigma$ increases further, the system starts to desynchronize due to the presence of those "runaway" oscillators.  Now there are two types of oscillators present in the system - phase-locked and runaway oscillators. The phase-locked oscillators always have a stable configuration, but the runaway oscillators may orient themselves in infinite number of possible ways. In fact, these oscillators bring homogeneity into the system at high $\sigma$ values. At any instant of time, the system remains in any one of its such possible states. Now, when a desynchronized state gets perturbed, it goes to one of its other possible states making the previous states locally unstable. So, despite the desynchronized state being globally stable, the states are locally unstable. Thus positive eigenvalues appear in the eigen-values spectra of the Jacobian in the region $\sigma \geq \sigma_c$. The number of such eigen-values increase with the increase of $\sigma$, as higher $\sigma$ increases the possibility of more locally unstable directions in the system, and finally it would not have any stable directions at very high values of $\sigma$. The onset of local instability of the de-synchronized solutions thus gives a measure of crossover noise-strength ($\sigma_c$).

\begin{figure}[]
\begin{subfigure}{.49\textwidth}
\hspace{-1.0 cm}
\includegraphics[scale=0.68]{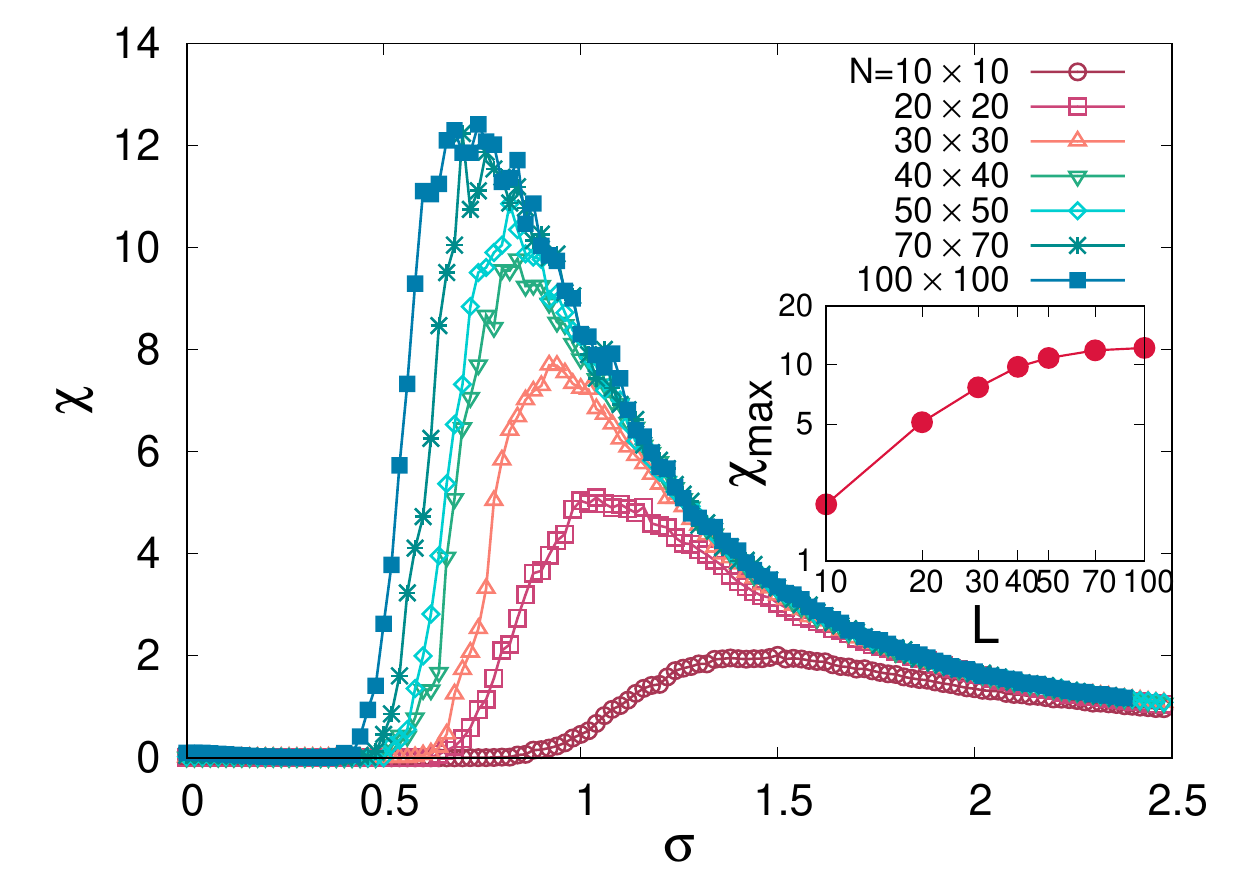}
\caption{}
\end{subfigure}
\begin{subfigure}{.49\textwidth}
\hspace*{-0.5 cm}
\includegraphics[scale=0.7]{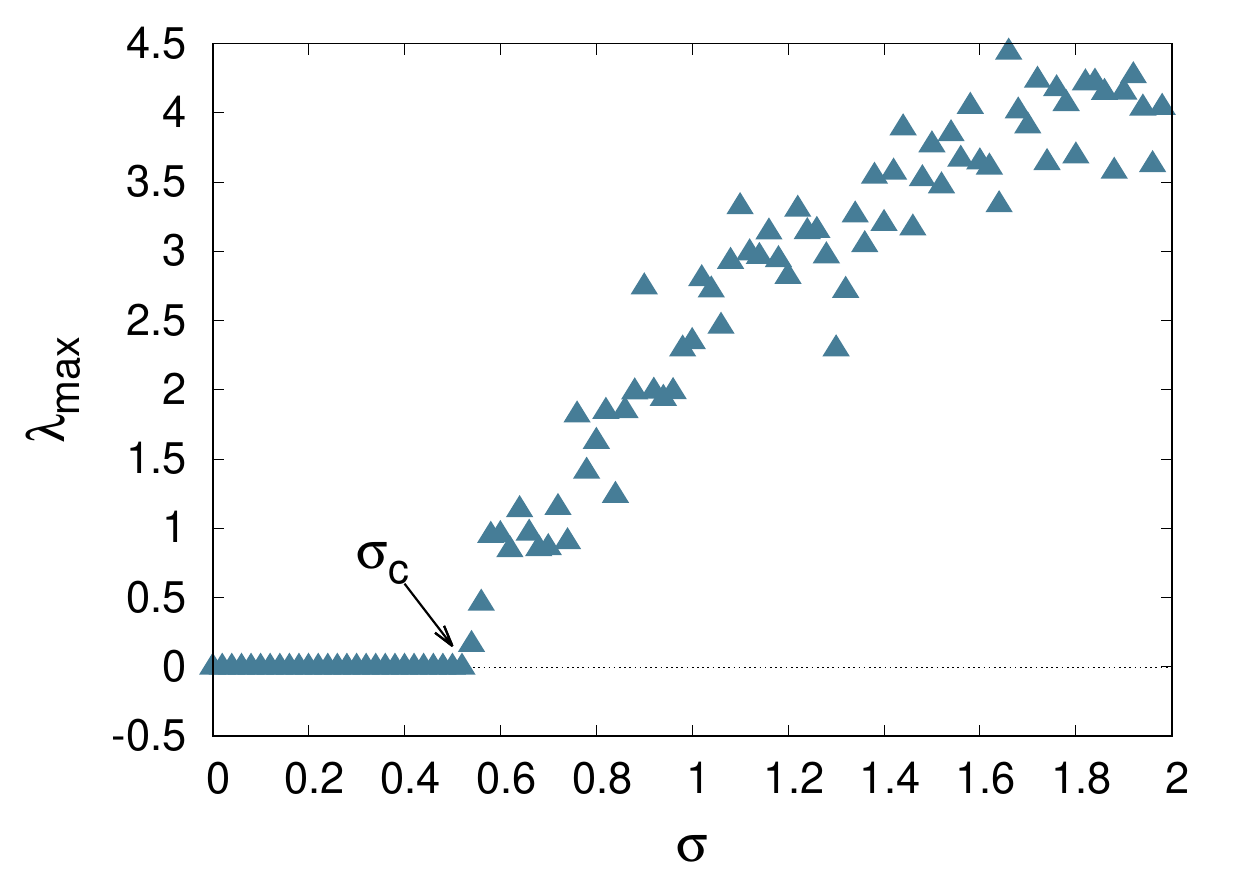}
\caption{}
\end{subfigure}\\
\begin{center}
\includegraphics[scale=0.7]{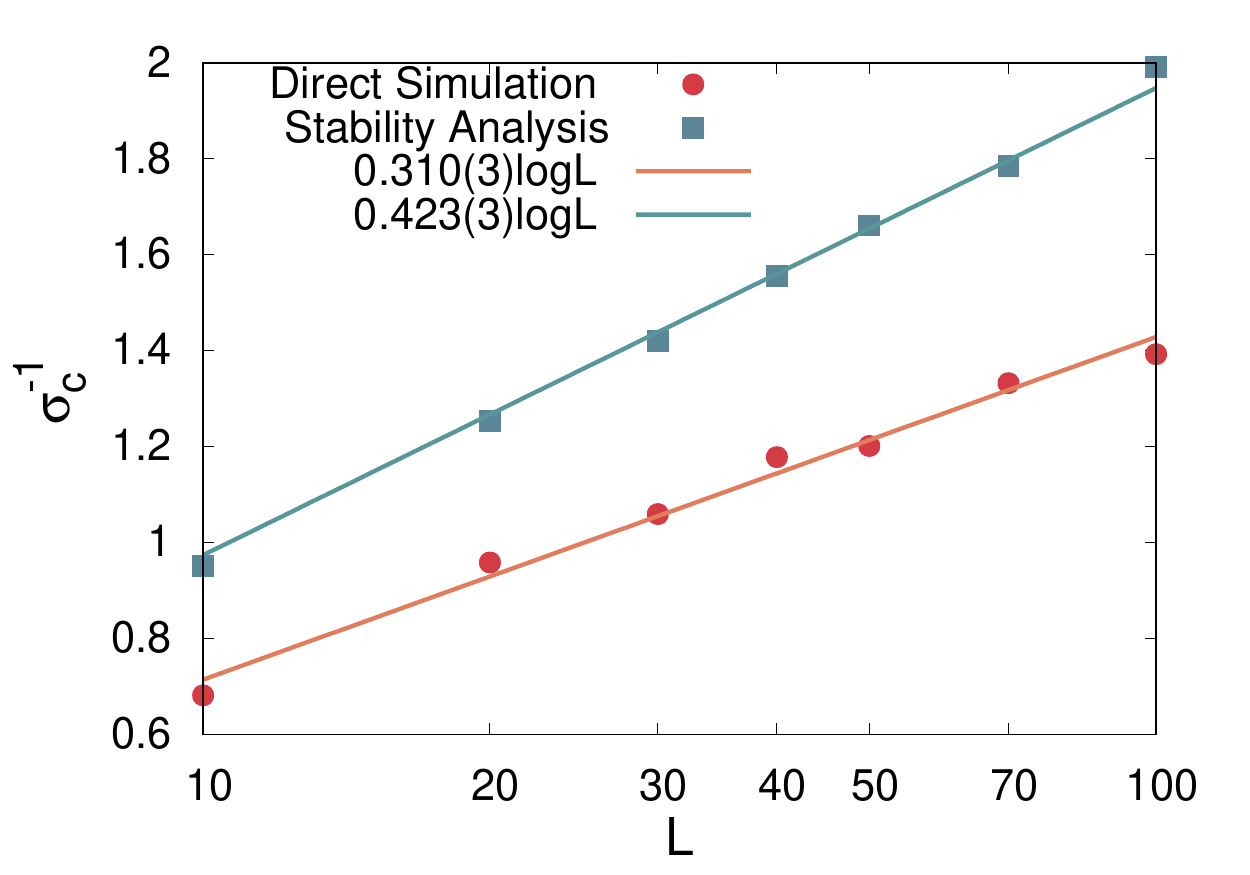}\\
(c)
\end{center}
\caption{(Color online) (a) Variation of dynamic fluctuations ($\chi$) with quenched noise-strength ($\sigma$) is shown for various system-sizes $N$ ($=L \times L$). Inset: $\chi_{\rm max}$ is plotted against linear system-size ($L$) on a logscale. The absence of power law behaviour indicates non-diverging (finite) correlation length, suggesting absence of any conventional phase transition in this system. (b) Largest eigen-value ($\lambda_{\rm max}$) of the Jacobian for a system-size $N=50 \times 50$ is plotted against noise-strength $\sigma$ for a particular realization of quenched disorder. The crossover noise-strength $\sigma_c$ is the point where $\lambda_{\rm max}$ becomes positive indicating the instability of the unsynchronized solution. (c) System-size dependency of the crossover noise-strength $\sigma_c$ is shown on a semi-log scale. Red circles denote the inverse $\sigma_c$ values calculated from direct simulation whereas the blue squares are the ones calculated from the linear stability analysis. A linear fit through the datapoints shows $\sigma_{c}^{-1} = 0.31 \log(L)$ (red line) for the direct simulation, whereas for linear stability analysis, $\sigma_{c}^{-1} = 0.42 \log(L)$ (green line).}
\label{fig:chi_vs_sigma_and_sigma_c_with_L_quenched_noise}
\end{figure}

For a particular system-size ($N=L\times L$), we calculate $\sigma_c (L)$ from the variation of the largest eigenvalue ($\lambda_{\rm max}$) with noise-strength $\sigma$, by locating the position of $\sigma$ where $\lambda_{\rm max}$ becomes positive, leaving the $x$-axis. The $\sigma_c(L)$ is averaged over 100 independent realization of quenched disorder. Figure  \ref{fig:chi_vs_sigma_and_sigma_c_with_L_quenched_noise}(b) shows the behaviour $\lambda_{\rm max}$ with $\sigma$ for a system of size $N=50 \times 50$ for one such realization of quenched disorder. The disordered averaged $\sigma_c (L)$ is plotted for different system-sizes $L$ and is shown in Figure \ref{fig:chi_vs_sigma_and_sigma_c_with_L_quenched_noise}(c) (squares in blue) on a semilog scale. A linear fit through the datapoints shows $\sigma_{c}^{-1} = 0.42 \log(L)$, whereas from direct simulation $\sigma_{c}^{-1} = 0.31 \log(L)$. Previous work on the crtical coupling obtained from phase-locking criterion in the system also supports this logarithmic scaling \cite{lee2010vortices}.

\clearpage

\section{Relaxation Dynamics}
\label{RD}

In this section we investigate how the system relaxes to equilibrium or non-equilibrium stationary states by observing the time evolution of the order parameter ($\rho$) at different noise-strengths for both kinds of noise.

\subsection{Oscillators with annealed noise}
\begin{figure}[]
\begin{subfigure}{.49\textwidth}
\hspace{-1.5 cm}
\includegraphics[scale=0.8]{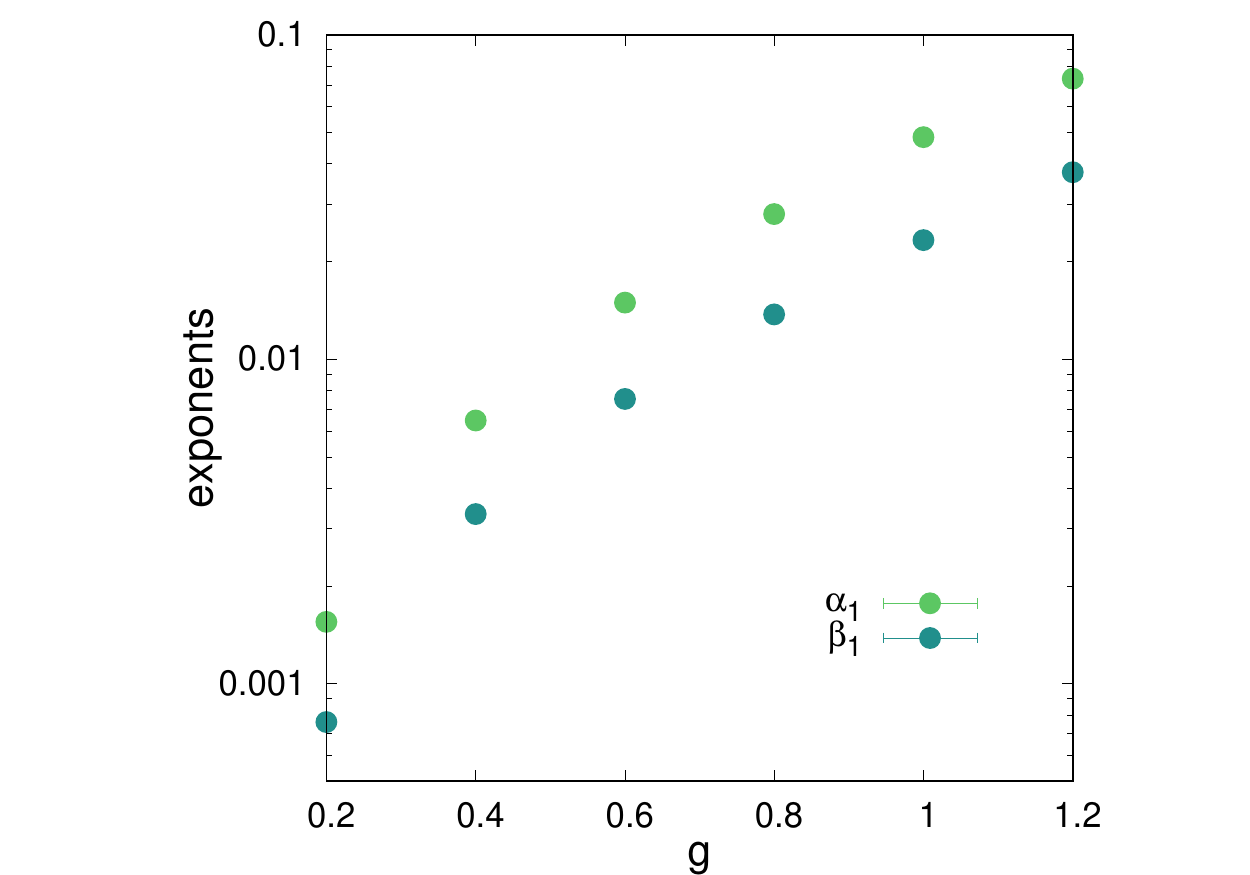}
\caption{}
\end{subfigure}
\begin{subfigure}{.49\textwidth}
\hspace*{-1.5 cm}
\includegraphics[scale=0.8]{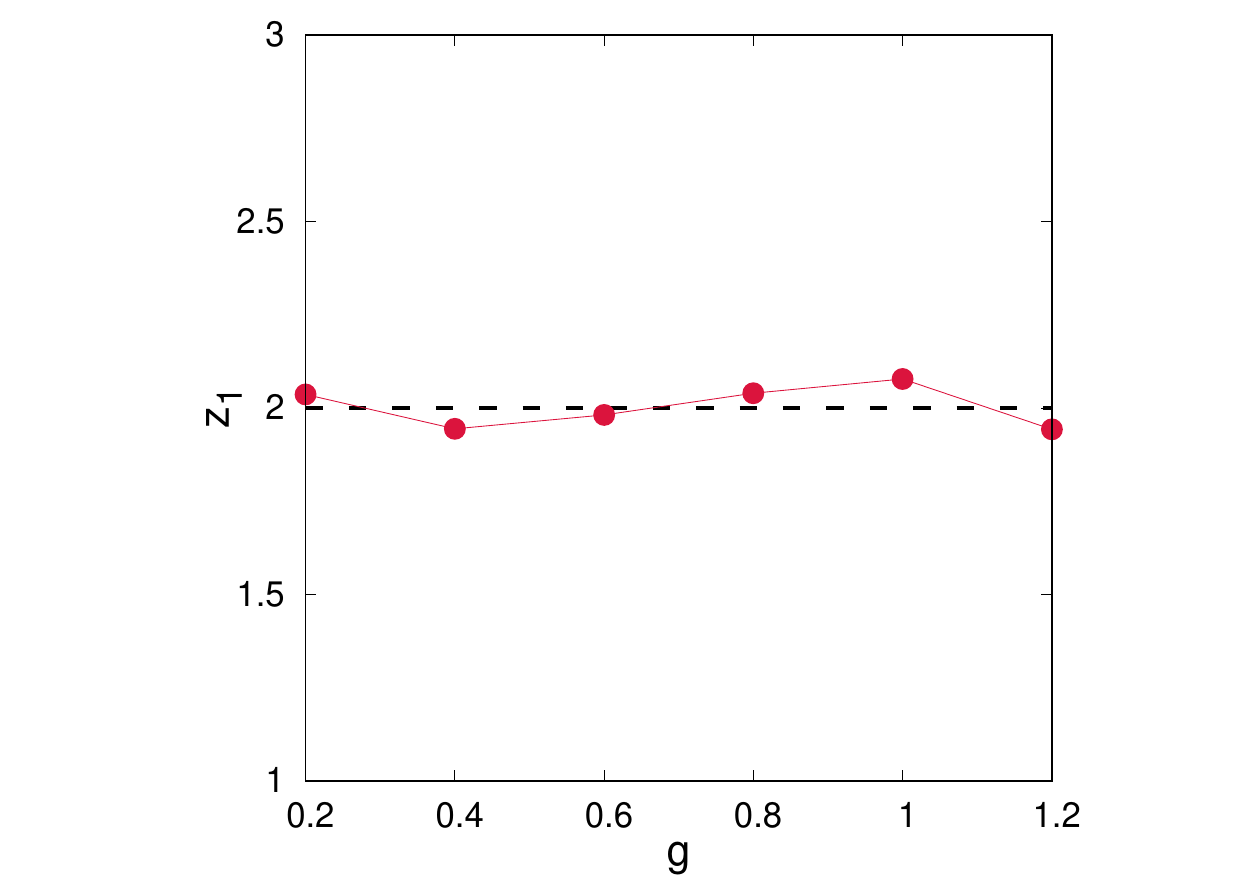}
\caption{}
\end{subfigure}
\caption{(Color online) (a) Temperature dependence of the exponents $\alpha_1, \beta_1$ in critically ordered phases is shown on a logscale. (b) Temperature dependence of the dynamic exponent ${\rm z}_1$ in critically ordered phases is shown. It shows that the dynamic exponents always remains ${\rm z}_1 \approx 2.0$ in this region.}
\label{fig:RD_g_temp_dependent_exponents}
\end{figure}

The system with annealed noise, starting from a synchronized state, relaxes to equilibrium differently in the two different phases. In the critically ordered phase, it relaxes to the equilibrium state algebraically. The power law exponent is found to be temperature dependent. Based on numerical simulations, we write the following scaling law for the order parameter as a function of system size ($L$) and time ($t$):
\begin{eqnarray}
\rho (L,t) = L^{-\alpha_{1}} \hat{\rho}(tL^{-z_1}),
\end{eqnarray}
where the scaling function $\hat{\rho}$ satisfies the following properties:
\begin{align}
\hat{\rho}(x) \sim  \left\{ \begin{array}{lr}{\rm const.}, \qquad & x \gg 1,\\
  x^{-{\alpha_{1}}/{z_1}},  \qquad  & x \ll 1.
\end{array}
\right.
\end{align} 
 So, in short-time regime ($tL^{-z_1} \ll 1 $), order parameter behaves as $\rho(L,t) \sim t^{-\beta_1} $ ($\beta_1 = {\alpha_1}/{z_1}$) and in long-time limit ($tL^{-z_1} \gg 1 $), in the stationary state, it scales as $ \rho_{\rm{st}} \sim L^{-\alpha_1}$. Here, $z_1$ is called the dynamic exponent.
Both the exponents, $\alpha_{1}$ and $\beta_{1}$ ($\beta_1$ is generally known as growth exponent in the language of surface growth model), vary continuously with temperature in such a way that the ratio ($z_{1}=\alpha_{1}/\beta_{1}$, dynamic exponent) always remains the same with the value $z_{1} \approx 2$. Figure \ref{fig:RD_g_temp_dependent_exponents} shows the variation of the exponents $\alpha_1,\beta_1$ (\ref{fig:RD_g_temp_dependent_exponents}(a)) and $z_1$ (\ref{fig:RD_g_temp_dependent_exponents}(b)) with temperature in this region.

We do not have proper argument in support of this, but atleast at low temperatures, it can be justified as follows. The system, in the low temperature regime, can be written as  
	\begin{eqnarray}
\frac{\rm \partial}{\partial t} \theta(\vec{\rm x},t) = \nabla^{2} \theta (\vec{\rm x},t) + g \xi (\vec{\rm x},t),
\end{eqnarray}
which is equivalent to the surface growth Edwards-Wilkinson (EW) model \cite{pal1999edwards,forrest1990hypercube}. The phase $\theta (\vec{\rm x}, t)$ is equivalent to the front height of the growing surface at position $\vec{\rm x}$ and time $t$. In general, the coupling constant which got absorbed during proper rescaling of time, acts as surface tension or, equivalently, here we assume the surface tension to be unity. The dynamic exponent for EW model in $2+1$ dimension is given by $z=2$ (Appendix \ref{appendix_EW}). So we expect in our system also the same value of the dynamic exponent which is consistent with the numerics. Surprisingly, this EW model can describe the relaxation dynamics not only at low temperatures, but also in the whole range of temperature over which critically ordered phase exists ($g \leq g_c$).

\begin{figure}[]
\begin{subfigure}{.49\textwidth}
\hspace{-1.5 cm}
\includegraphics[scale=0.4]{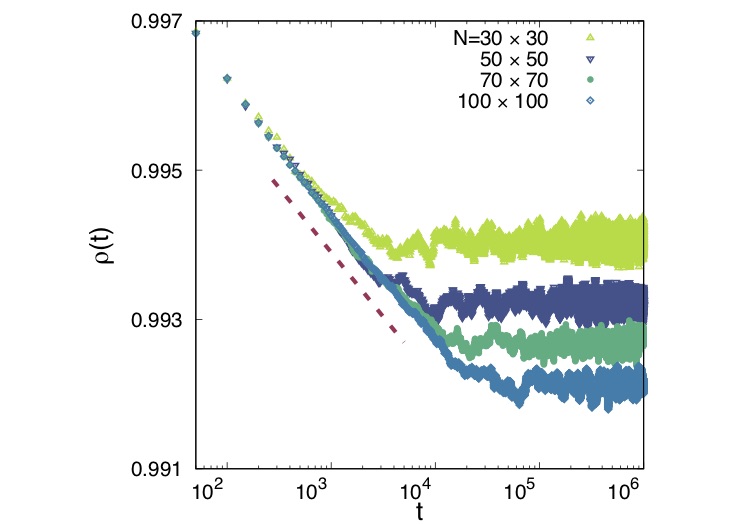}
\caption{}
\end{subfigure}
\begin{subfigure}{.49\textwidth}
\hspace*{-1.0 cm}
\includegraphics[scale=0.8]{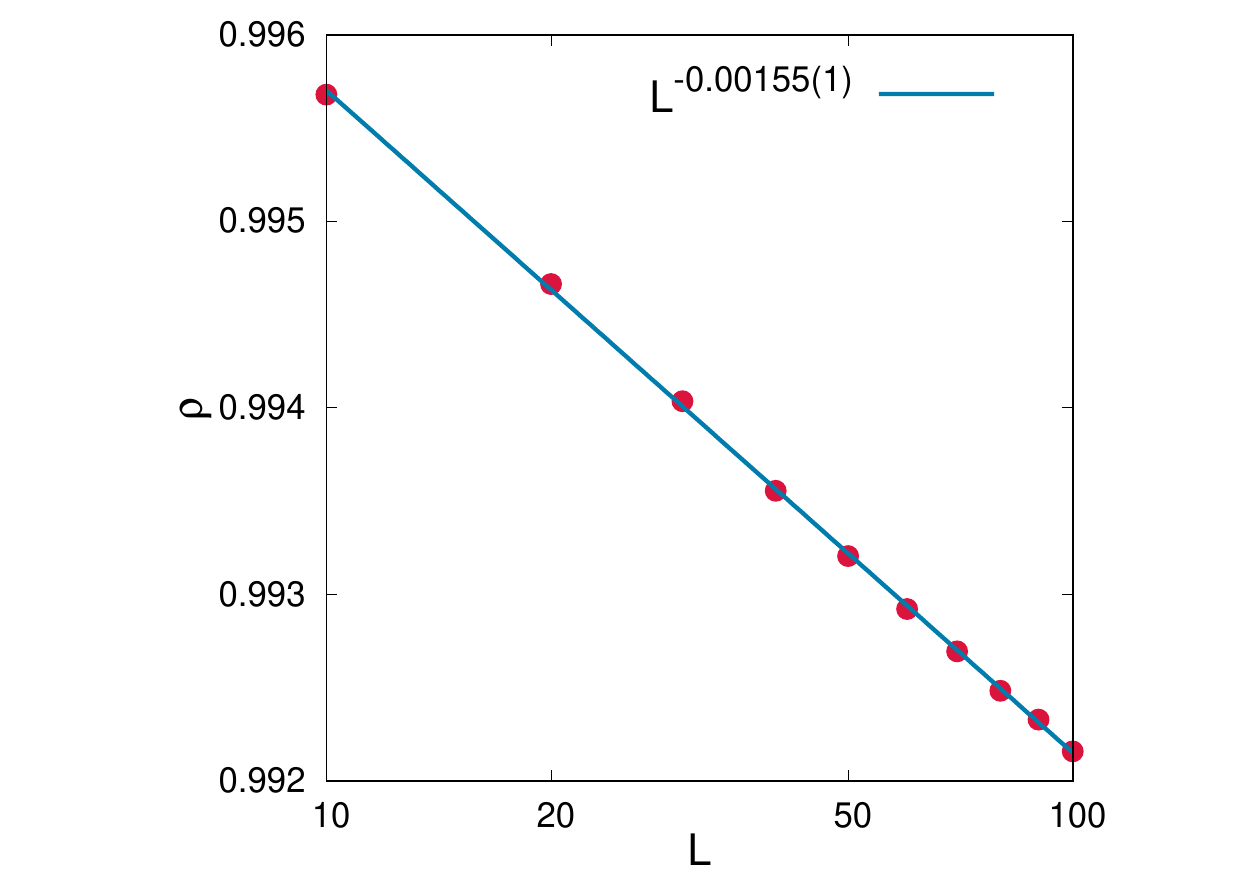}
\caption{}
\end{subfigure}
\begin{subfigure}{.49\textwidth}
\hspace*{-1.5 cm}
\includegraphics[scale=0.39]{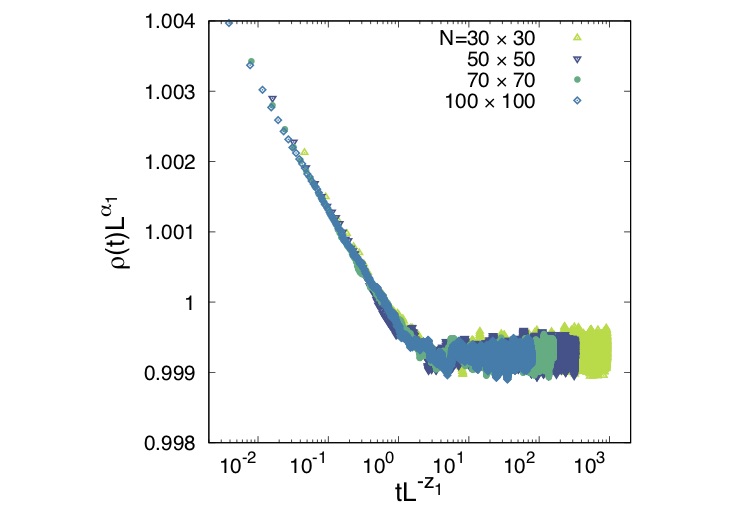}
\caption{}
\end{subfigure}
\begin{subfigure}{.49\textwidth}
\hspace*{-0.75 cm}
\includegraphics[scale=0.38]{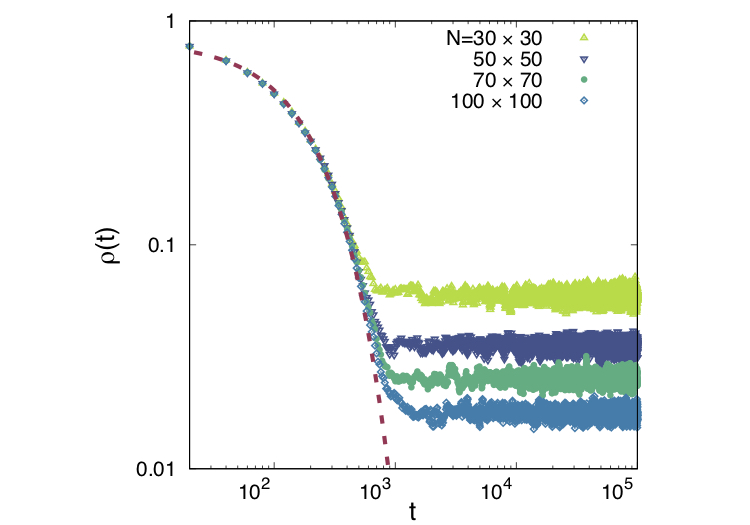}
\caption{}
\end{subfigure}
\caption{(Color online) (a) Time evolution of $\rho$ is shown on a logscale for different system sizes $L$ at $g=0.2$. (b) Scaling plot of $\rho$ in the steady state for for various $L$ is shown on a logscale. (c) Scaling of $\rho$ on a logscale is shown. A good collapse of the data is achieved for the value of $\alpha_1$ and $z_1$ estimated from (a) and (b). (d) Exponential relaxation at a particular temperature $g=2.0$ in the high temperature region is shown.}
\label{fig:RD_g_scaling}
\end{figure}

Figure \ref{fig:RD_g_scaling}(a) shows, on a logscale, how $\rho$ evolves in time for various $L$ at $g=0.2$. The linear portions are fitted with a straight line giving a slope $\beta_1 = 0.00075(1)$. In the large-time limits, when the system is in saturation, $\rho$ is calculated and plotted in Figure \ref{fig:RD_g_scaling}(b). A linear fitting, on the log-log plot, through the datapoints yields $\alpha_1 = 0.00155(1)$, which gives an estimation of $z_1 =2.07$. The scaling plot of the $\rho$  with time ($t$) for $L$ is shown in Figure \ref{fig:RD_g_scaling} on a log-scale. A good collapse of the data is evident for the choice of parameters estimated above. The value of the dynamic exponent ($z_1=2.07)$ estimated from the Kuramoto model agrees well with that ($z=2$) of EW model implying that the $2D$ Kuramorto model of identical oscillators belongs to the EW universality class in the region $g \leq g_c$.
 
In the disordered phase ($g > g_c$), the system decays to the equilibrium state exponentially fast with an exponent ($\lambda \approx 0.5$), where the time has been taken in natural time unit.
Figure \ref{fig:RD_g_scaling}(d) shows the log-log plot of exponential decay of the order parameter for various $L$ at $g=2.0$.

\subsection{Oscillators with quenched noise}
In this section, we study relaxation dynamics of the global phase order parameter ($\rho$) of the system under the influence of quenched noise at different noise-strengths. 
The system, in both the synchronized and desynchronized states, relaxes to stationary state exponentially.

\begin{figure}[]
\begin{subfigure}{.49\textwidth}
\hspace{-1.5 cm}
\includegraphics[scale=0.4]{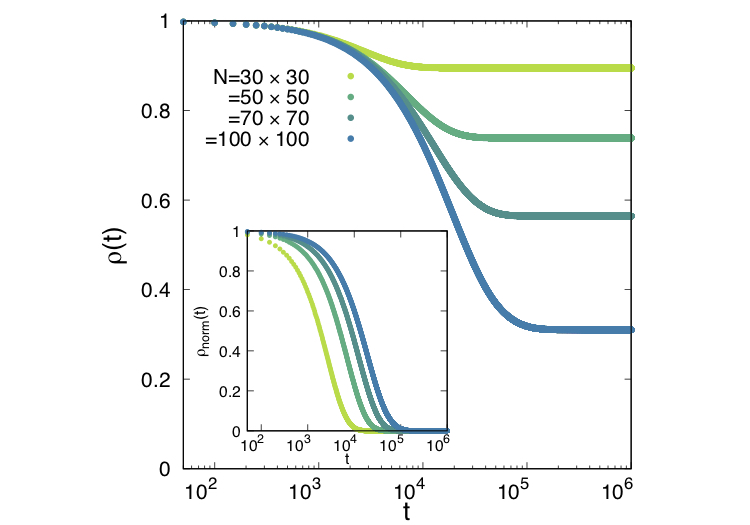}
\caption{}
\end{subfigure}
\begin{subfigure}{.49\textwidth}
\hspace*{-1.5 cm}
\includegraphics[scale=0.83]{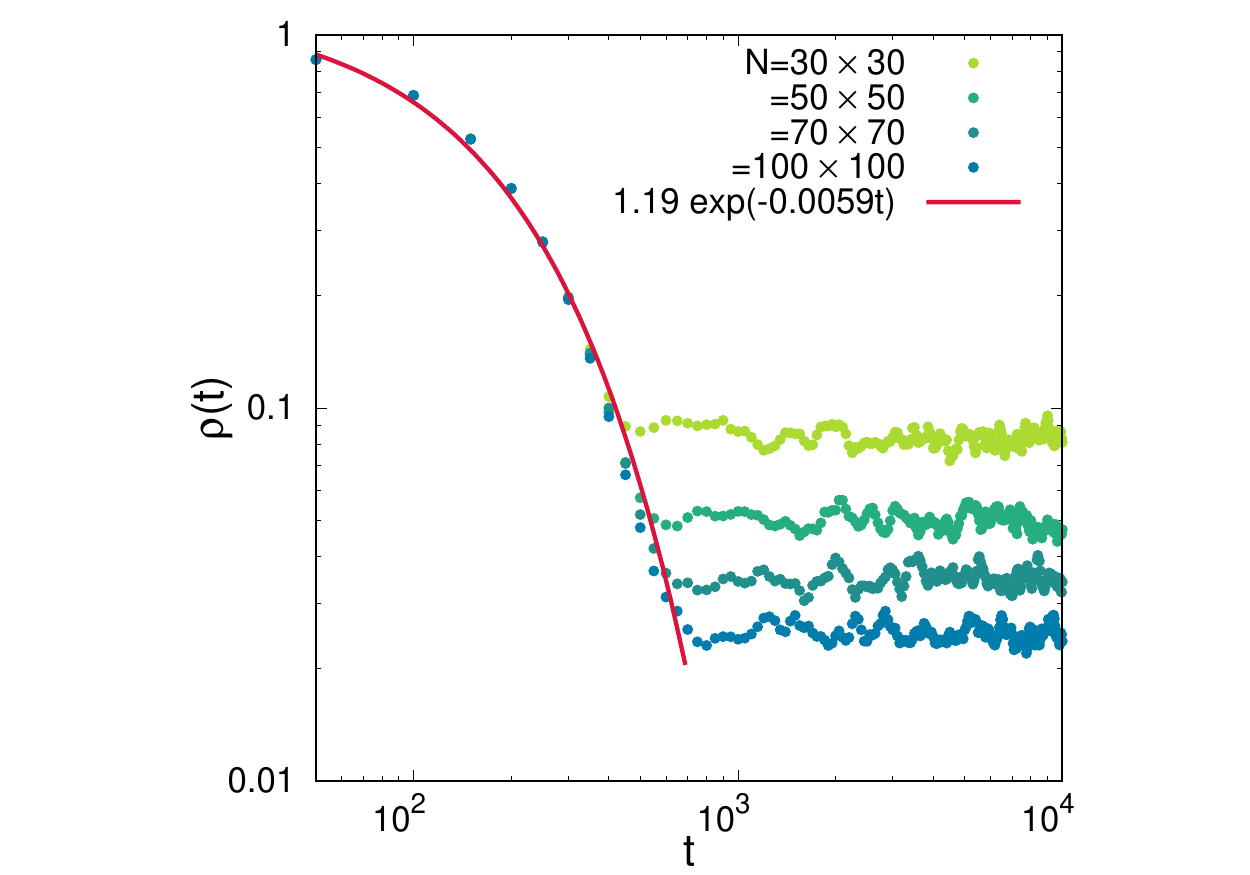}
\caption{}
\end{subfigure}
\caption{(Color online) (a) Time evolution of the Global order parameter at $\sigma = 0.2$ for various system-sizes is shown on a semilog scale. Inset: Time evolution of corresponding $\rho_{\rm norm}$ at the same parameter value is shown. (b) Log-log plot of time evolution of the Global order parameter at $\sigma = 2.0$ for various $L$. An exponential fit through the datapoints yields the exponent $\lambda \approx 0.59$, where the time has been considered in natural time unit.}
\label{fig:RD_ord_para}
\end{figure}

In such a scenario, we compute the relaxation time in the synchronized regime ( $\sigma \leq \sigma_{c}(L)$) which gives an estimation of the time required to transfer information through the network. The average relaxation time is defined by \cite{kim1997anomalous,son2008relaxation}
\begin{equation}
\tau_{\rm av} = \int_{0}^{\infty} {\rm dt'} \rho_{\rm norm}(t') ,
\end{equation}
where, the normalized order parameter ($\rho_{\rm norm}$) is defined as follows:
\begin{equation}
\rho_{\rm norm}(t)=\frac{\rho(t) - \rho_{\rm st}}{\rho(0) - \rho_{\rm st}}.
\end{equation}
So initially (t=0), $\rho_{\rm norm} = 1$ and at saturation ($t \to \infty$), it becomes 0. Time evolution of the Global order parameter at a particular small $\sigma = 0.2 (< \sigma_c(L)$ for all $L$ values considered here) for various system-sizes is shown in Figure \ref{fig:RD_ord_para}(a) on a semi-logscale. The time evolution of the corresponding normalized order parameters is shown in the inset of Figure \ref{fig:RD_ord_para}(a). This shows the exponential relaxation. The dynamics at very high noise-strength ($\sigma \gg \sigma_c(L)$) is also exponential, as shown in the Figure \ref{fig:RD_ord_para}(b). Here, the time-evolution of $\rho$ is plotted on a logscale for various values of $L$ at $\sigma =2.0$, which fit well with an exponential function with an exponent $\approx 0.59$ (the time being considered in natural time unit). 

\begin{figure}[]
\begin{subfigure}{.49\textwidth}
\hspace{-1.85 cm}
\includegraphics[scale=0.8]{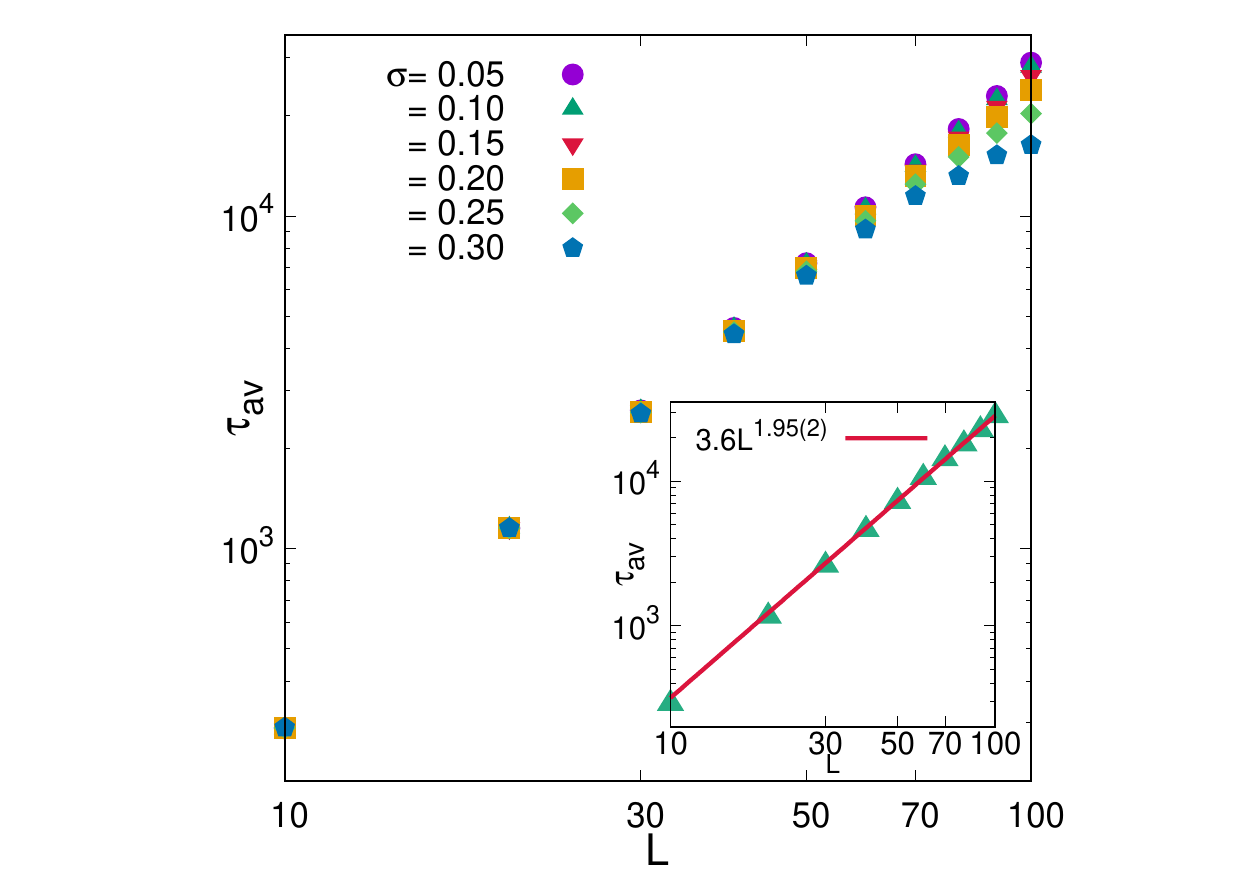}
\caption*{Left}
\end{subfigure}
\begin{subfigure}{.49\textwidth}
\vspace*{+0.3 cm}
\hspace*{-1.0 cm}
\includegraphics[scale=0.8]{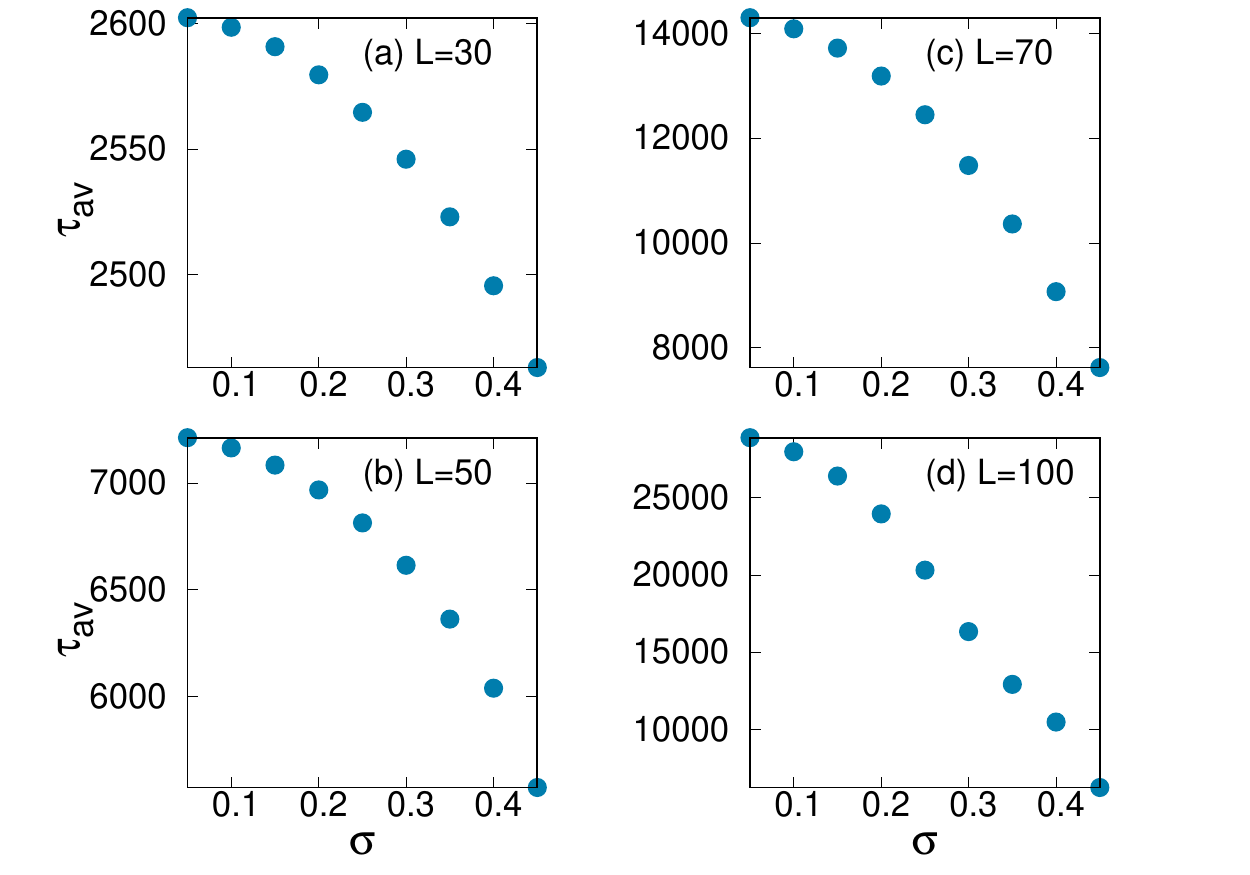}
\caption*{Right}
\end{subfigure}
\caption{(Color online) Left: Plot of ${\tau}_{\rm av}$ with $L$ for various values of $\sigma$ is shown on a logscale. Inset: Log-log plot of ${\tau}_{\rm av}$ with $L$ for $\sigma=0.1$. A linear fiiting through the datapoints shows the power law scaling with the dynamic exponent $z=1.95(2)$.
Right: Variation of ${\tau}_{\rm av}$ with $\sigma$ for $L=30,50,70$ and $100$ is shown in subfigure (a),(b),(c) and (d) respectively.}
\label{fig:RD_tau_av_sigma_N}
\end{figure}

 In the weak noise-strength regime (synchronized regime), we investigate the behaviour of the average relaxation time ($\tau_{\rm av}$) with the noise-strength ($\sigma)$ and the system-size($L$). We compute $\tau_{\rm av}$ for various values of $L$ at different noise-strength $\sigma$ in the synchronized regime and is shown in Figure \ref{fig:RD_tau_av_sigma_N}(Left) on a logscale.
For very small $\sigma$ value, $\tau_{\rm av}$ follows a power-law behaviour with the linear system-size ($\tau_{\rm av} \sim L^{z}, z$ being dynamic exponent) as shown in the the inset of figure \ref{fig:RD_tau_av_sigma_N}(Left), where $\tau_{\rm av}$ scales with $L$ for a particular $\sigma = 0.1$. The power law fitting yields the dynamic exponent z=1.95. As $\sigma$ increases, the system of large $L$ deviates from its linear behaviour on the log-log plot. Next, we study how the average relaxation time, for a particular system-size($L$), varies with $\sigma$ in synchronized regime $\sigma \leq \sigma_c(L)$. It shows that, $\tau_{\rm av}$ decreases with the increase of $\sigma$. Figure \ref{fig:RD_tau_av_sigma_N}(Right) depicts this scenario. 

\section{Discussion and Conclusion}
In summary, we have systematically explored the phase synchronization in a system of locally coupled Kuramoto oscillators arranged on a $2D$ square lattice with annealed and quenched types of disorder. In the bare Kuramoto model with annealed noise, we unveil $KT$-type phase transition in the thermodynamic limit, as observed in $2D$ $XY$ model, via numerical investigations on finite systems of various sizes. We obtain the critical temperature as well as the critical exponents associated with the transition using finite-size scaling (FSS) theory. 
In case of quenched noise, 
we re-establish that phase-locking is not possible in this system in the thermodynamic limit \cite{strogatz1988phase,hong2005collective}. Here, we were able to analyze the crossover in the system of finite size via linear stability analysis of the stationary state solutions. We assume that the local instability of the unsynchronized solution occurs due the presence of "runaway" oscillators in the system. Based on this simple idea, we were also able to obtain numerically the system-size dependent crossover noise-strength from the onset of local instability of the unsynchronized solution, which is in well agreement with the earlier work \cite{lee2010vortices}.
    
We also observe that the system relaxes to equilibrium differently for these two types of noise. In case of annealed noise, the system is found to exhibit algebraic relaxation in the critically ordered phase which is described by the phenomenological Edward-Wilkinson model of growing surface. We recover the dynamic exponent $z=2$ for this case, suggesting that the model with annealed noise belongs to the $EW$ universality class. But, in the disordered phase ($g > g_c$), it shows an exponential decay. The system with quenched noise, as opposed to annealed one, always relaxes to equilibrium exponentially.

So far, we have studied the system with the synchronized initial conditions i.e. the phases of all the oscillators were set to zero. The random initial conditions introduce topological defects in the phase-field of the oscillators. The next step would be to study the role of these defects in the synchronization phenomena. Finally, it would also be interesting to study the system in presence of these annealed and quenched types of noise together, especially to investigate whether one can destroy the critically ordered phase by introducing quenched disorder into the system. But, in such cases, the system dynamics would no longer be an equilibrium phenomenon. One should look for if any non-equilibrium phase transition is possible or not in such a system. 

\clearpage
 
\section*{Acknowledgments}
I thank Neelima Gupte for introducing me to the field of synchronization in many-body interaction systems. I acknowledge very useful discussions on KT transition with Mustansir Barma. I am grateful to Shamik Gupta for fruitful discussions and suggestions on the manuscript. I also thank HPCE, IIT Madras for providing me with high performance computing facilities in VIRGO Super cluster. 

\section*{Appendix}
\appendix
\section{Equivalence between the Kuramoto model and the Classical XY model}
\label{appendix_KM_XY_equivalence}

The model we considered has a connection with a classical Hamiltonian system, the $XY$ model, another simple paradigmatic model to study magnetic systems. We start with the Hamiltonian of classical $XY$ model on a lattice where the spins interact with their nearest neighbours only, 
\begin{equation}
   H = \sum_{i}  \frac{p_{i}^{2}}{2m} - K \sum_{j \in nn_{i}}  \vec{s}_{i}.\vec{s}_{j}
     = \sum_{i}  \frac{p_{i}^{2}}{2m} - K \sum_{j \in nn_{i}}  \cos(\theta_i-\theta_j),
\end{equation}
where \{$\theta_i$, $p_i$\} are the canonical conjugate dynamical variables associated with the $i$-th spin. Here, $nn$ represents that the sum is over nearest neighbours only and $K(>0)$ is the strength of interaction. In fact, the conjugate momentum, $p_i \equiv mv_i$, where $m$ is the mass and $v_i$ is its velocity.

Thus the Hamilton's equations of motion read as follows:
\begin{align}
    \frac{{\rm d} \theta_{i}}{\rm dt} &= \frac{p_i}{m} \label{Eq:velocity_term},\\
    m\frac{{\rm d} v_{i}}{\rm dt} &= - K \sum_{j \in nn_{i}} \sin(\theta_i -\theta_j). \label{Eq:force_term}
\end{align}

Now, we consider a situation where the system is in contact with a heat bath and it experiences an additional external drive which acts as a torque $\sigma \omega_i$ on individual spins. The strength of the external torque is represented by $\sigma(>0)$. So, if $\gamma$ be the damping constant, the dynamics of spin (Equation \ref{Eq:force_term}) gets modified as follows, namely the Langevin equation:
\begin{equation}
    m\frac{{\rm d} v_{i}}{\rm dt} = -\gamma v_i + \gamma \sigma \omega_i - K \sum_{j \in nn_{i}} \sin(\theta_i -\theta_j) + \sqrt{\Gamma} \eta_{i}(t),
    \label{Eq:modified_force_term}
\end{equation}
where $\eta_{i}(t)$, the Gaussian white noise, represents the thermal fluctuation in the system due to the presence of the heat bath and is characterized by
\begin{equation}
\langle \eta_{i}(t) \rangle = 0    \qquad\mbox{and}\qquad  \langle \eta_{i}(t)\eta_{j}({t'}) \rangle = \Gamma \delta_{ij} \delta(t - {t}^{\prime}),
\end{equation}
where, $\langle \cdot \rangle$ denotes averaging over noise realizations, and $\Gamma$ is the strength of the Gaussian noise. From fluctuation dissipation theorem, $\Gamma = 2 m \gamma k_{B} T$, where $T$ is the temperature of the heat bath and $k_{B}$ is the Boltzmann constant. Here we assume that the heat bath, having enormously large degrees of freedom as compared to the that of the system of our interest, does not get affected significantly by the presence of it.

We want to put an emphasis for the reader on the source of damping over here. The damping originates here due to the presence of the medium (heat bath) where the system is immersed in. Recent studies show that this system (classical $1D$ $XY$ model), with angular-momentum conserving Langevin dynamics, can mimic the non-Newtonian flow regimes under certain conditions \cite{evans2015classical}. The flow is considered as collections of these individual rotors. In such case, the damping originates due to the relative motion between the rotors.
We must keep in mind that the model itself, in the later case, represents the medium. The damping we considered in Equation \ref{Eq:modified_force_term} is different from the later one which is beyond the scope of our study.

The system dynamics is now governed by the Equation \ref{Eq:velocity_term} along with Equation \ref{Eq:modified_force_term}, from which the overdamped dynamics can be obtained in the limit $m/\gamma \ll 1$. As before, by proper rescaling of time and other parameters, we obtain the overdamped dynamics as
\begin{equation}
    \frac{{\rm d} \theta_{i}}{\rm d\tilde{t}} =  \tilde{\sigma} \omega_{i} + \sum_{j \in nn_{i}}  \sin(\theta_{j}- \theta_{i}) + \tilde{g} \xi_{i}(\tilde{t}),
    \label{Eq:xy_overdamped}
\end{equation}
where, 
\begin{equation*}
    \tilde{t} \equiv t \frac{K}{\gamma},\qquad \tilde{\sigma} \equiv \sigma \frac{\gamma}{K},\qquad \xi_{i}(\tilde{t}) \equiv \eta_{i}(t) \frac{\sqrt{\gamma}}{K} \qquad\mbox{and}\qquad \tilde{g} \equiv \sqrt{\frac{\Gamma}{K}}.
\end{equation*}
The spin dynamics (Equation \ref{Eq:xy_overdamped}) resembles the evolution equation for the Kuramoto model given by Equation \ref{govn_eqn_3}.
Thus the Kuramoto dynamics under study (Equation \ref{govn_eqn_4}) is equivalent to the overdamped dynamics of the classical $XY$ model with nearest neighbour interactions in contact with a heat reservoir and in presence of external drives. The external drives in the form of quenched disorder are equivalent to the intrinsic frequencies of the Kuramoto oscillators. This external torque drives the system to non-equilibrium stationary state. As long as, there is external drive present in the system ($\sigma \neq 0$), this is no longer a Hamiltonian system. Of course, we recover the Hamilton's equations of motion in the limit $\sigma=g \to 0$.

\section{KT Transition in the 2D XY model}
\label{appendix_KT}
According to Mermin-Wagner theorem, $2D$ $XY$ model does not exhibit usual kind of phase transition in the thermodynamic limit \cite{mermin1966absence}. But, this system exhibits an unusual kind of phase transition, called topological phase transition (also called $KT$ transition, after the name of Kosterlitz and Thouless), where the system makes a transition from quasi-long range order (critically ordered phase) to disorder \cite{kosterlitz1973ordering,kosterlitz1974critical}. In general a phase transition is characterized by non-zero order parameter in ordered phase and zero in the disordered phase in the thermodynamic limit. But in this case, it remains zero at all finite values of temperatures. Thus, there is no macroscopic ordering in the system even at low temperatures. In terms of fluctuations, it diverges in this region in the thermodynamic limit. For a continuous phase transition, there exists a critical temperature where the system shows critically ordered phase with algebraic decay of correlations and diverging correlation length in the thermodynamic limit. Now, in $2D$ $XY$ model, similar kind of critically ordered phase exists at all temperatures from zero to the critical temperature ($T_{c}$). Beyond this region ($T > T_{c}$), the system displays exponential decay of correlation. 
For $2D$ $XY$ model the critical temperature is $T_c = \pi/2$ in units of $J/k_{\rm B}$ ($J$ being the nearest neighbour coupling strength). 

For $KT$-transition in $2D$ $XY$ model, as $T \to T_c$, the correlation length $\xi$ and the susceptibility $\chi$ diverge according to the asymptotic laws \cite{kosterlitz1974critical}
\begin{eqnarray}
\xi &\sim \exp(C\epsilon^{-1/2}),  \qquad & \epsilon > 0 \label{Eq:chi_xy_KT},\\ 
	&= \infty,  \qquad & \epsilon \leq 0.
\end{eqnarray}
where $\epsilon = T-T_{c}$ and $C \approx 1.5$, and
\begin{eqnarray}
\chi &\sim \xi^{2-\eta},  \qquad & \epsilon > 0,\\ 
	&= \infty,  \qquad & \epsilon \leq 0.
\end{eqnarray}
where the exponent $\eta = 1/4$. So the correlation length $\xi$ falls faster than any power of $\epsilon$. As $\epsilon \to 0^{+}$, it diverges i.e. becomes infinite and this remains so for all $\epsilon \leq 0$.

The mechanism behind this phase transition is the following. There are always vortices and spin-waves present in the system. In the region $T \leq T_c$, the vortices are bound in pairs with total vorticity zero and spin-wave excitations are the dominant ones. These spin-wave excitations are responsible for destroying long-range order in the system. But above $T_c$, the vortices become unbound and they are now free to move to the surface under the influence of arbitrarily weak applied field, thereby causing a phase transition.

\section{Edwards-Wilkinson Model}
\label{appendix_EW}
The Edwards-Wilkinson (EW) model which describes surface fluctuation in a non-equilibrium system is given by \cite{pal1999edwards,forrest1990hypercube} 
\begin{eqnarray}
\frac{\rm \partial}{\partial t} \theta(\vec{\rm x},t) = \nu \nabla^{2} \theta (\vec{\rm x},t) + \xi (\vec{\rm x},t),
\label{govn_eqn_7}
\end{eqnarray}
 where, $\theta (\vec{\rm x}, t)$ is the front height of the growing surface at position $\vec{\rm x}$ and time $t$, $\nu$ is the surface tension and $\xi (\vec{\rm x},t)$ is the stochastic contribution to the surface fluctuations. 

In the context of surface growth model, the interfacial width (roughness of the surface) $w$ is defined as a function of linear system-size ($L$) and time ($t$) as
\begin{eqnarray}
w^{2}(L,t) = \left \langle \overline{\theta^{2}(\vec{\rm x},t)} - {\overline{\theta(\vec{\rm x},t)}^{2}}\right \rangle.
\end{eqnarray}
where $ \overline{\cdot}$ denotes the spatial average and $\langle \cdot \rangle$ denotes the average over noise realizations.
 
 In general, for arbitrary spatial dimension $d$, in short-time limits, the interfacial width scales as $w^{2} \sim t^{\beta}$ and in large-time limits, when the system is in equilibrium, it behaves as $w^{2} \sim L^{\alpha}$ where $\alpha$ and $\beta$ are called the roughness exponent and the growth exponent respectively. The lower and upper critical dimensions for EW class are $d^{l}_{c} = 0$ and $d^{u}_{c} = 2$, respectively. For the EW model in $2+1$ dimension, for sufficiently long time and large substrate size, the interfacial width follows the scaling law
 \begin{eqnarray}
 w^{2}(L,t) = A \ln \left [ L \hat{w} (tL^{-z}) \right ]
 \end{eqnarray}
 where the scaling function $\hat{w}$ satisfy the following properties:
 
\begin{align}
\hat{w}(x) \sim  \left\{ \begin{array}{lr}{\rm const.}, \qquad & x \gg 1,\\
 x^{\beta},  \qquad  & x \ll 1.
\end{array}
\right.
\end{align} 


 So, in short-time regime ($tL^{-z} \ll 1 $), interfacial width behaves as $w^{2}(L,t) \sim A \beta \ln t $ and in long-time limit ($tL^{-z} \gg 1 $), in the stationary state, it scales as $ w_{\rm{st}}^{2} \sim A \ln L + \rm{constant} $. $z$ is called the dynamic exponent and it is related to $\beta$ as $z\beta = 1$.
 
\begin{figure}[]
\begin{subfigure}{.49\textwidth}
\hspace{-1.5 cm}
\includegraphics[scale=0.4]{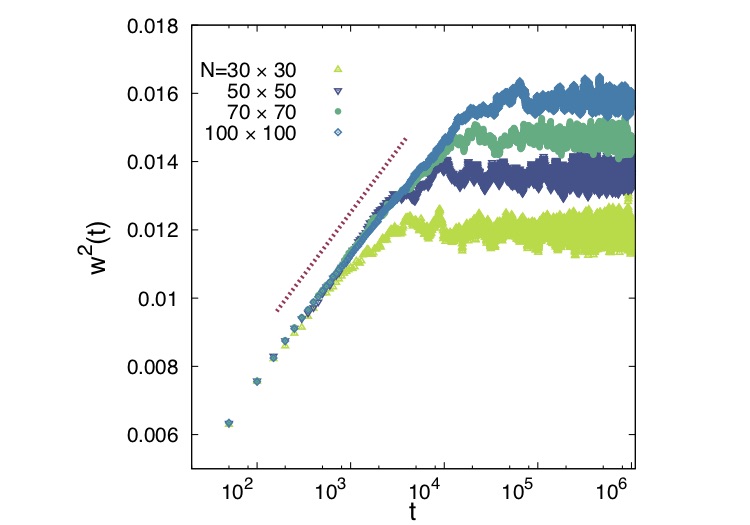}
\caption{}
\end{subfigure}
\begin{subfigure}{.49\textwidth}
\hspace*{-1.5 cm}
\includegraphics[scale=0.8]{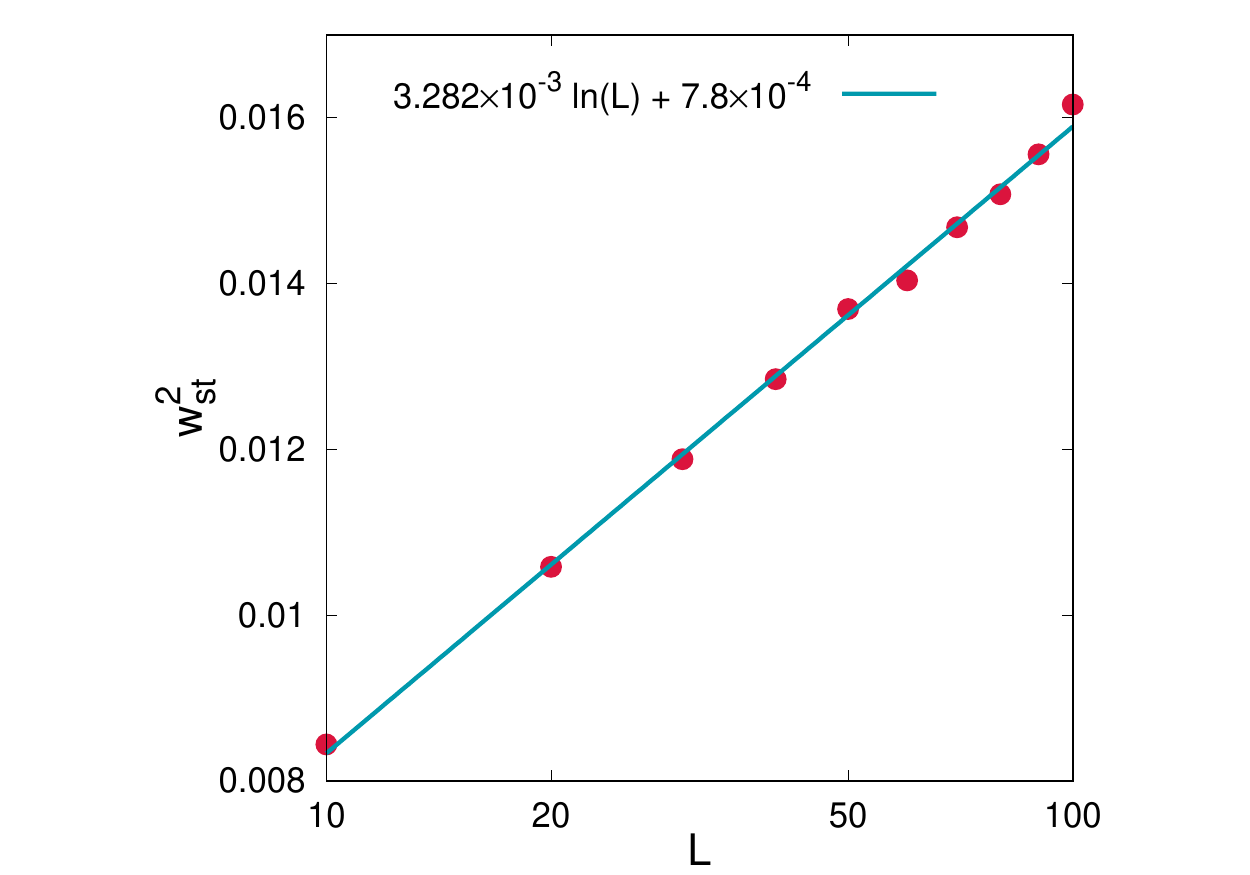}
\caption{}
\end{subfigure}\\
\begin{center}
\includegraphics[scale=0.4]{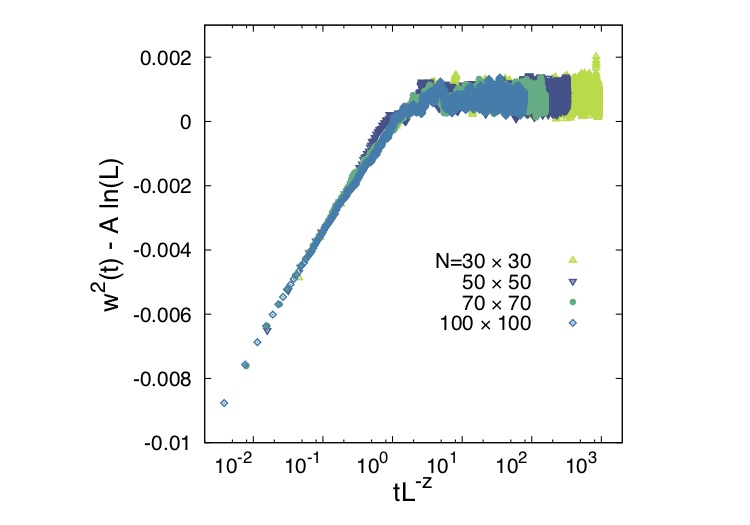}\\
(c)
\end{center}
\caption{(Color online) (a) Interfacial width in short time is plotted against time on a semilog scale for different system sizes. These datapoints are obtained  by averaging over 100 different realizations of noise and initial distribution of the oscillators were chosen to be zero. The data shows a good fitting with the theoretical logarithmic function predicted from the EW model. (b) Variation of interfacial width in the stationary state at $g=0.2$ for different system sizes on a logscale is shown. The data shows a good fitting with the theoretical logarithmic function predicted from the EW model. (c) Scaling of interfacial width on a semilog scale is shown. A good collapse of the data is achieved for the obtained value of the dynamic exponent ($z=2.0086$). This value is close to that of the theoretical prediction for EW model ($z=2$). It shows that, in the critically ordered phase, the phase-variance in the KM model behaves as interfacial width in surface growth EW model.}
\label{fig:EW_model}
\end{figure}

 From the stationary distribution of the oscillators with annealed noise in the linear regime, using the above scaling laws, we obtain a relationship between the order parameter and the phase variance as 
 \begin{eqnarray}
 \rho = \langle e^{{\rm i} (\theta -\bar{\theta})} \rangle=\exp (- w^{2}/2),
 \end{eqnarray}
 which validates the algebraic decay of the relaxation. We now calculate the dynamic exponent $z$ using the above scaling law.
 
Figure \ref{fig:EW_model}(a) shows how the interfacial width for Kuramoto model (fluctuation of the phases over the entire lattice) grows in time (starting from a flat surface initially) for various substrate sizes at $g=0.2$. The linear portions are fitted with a straight line giving a slope $A\beta = 0.001634(4)$. In the large-time limits, when the system is in saturation, $w^{2}$ is calculated for different substrate sizes and are plotted in Figure \ref{fig:EW_model}(b). A logarithmic fitting through the datapoints yields $A = 0.003282(59)$, which gives an estimation of $\beta =0.49787$. From the identity $z\beta =1$, $z$ is estimated as $z=2.0086$. The scaling plot of the mean-square width with time ($t$) for substrate sizes $L$ is shown in Figure \ref{fig:EW_model}(c), which shows a good collapse of the data for the choice of parameters estimated above. The value of the dynamic exponent ($z=2.0086)$ estimated from linearized version of the Kuramoto model agrees well with that ($z=2$) of EW model implying that the $2D$ Kuramorto model of identical oscillators belongs to the EW universality class in the region $g \leq g_c$.

\clearpage

\bibliography{stochastic_KM_2D_mrinal}

\end{document}